\documentclass[aps,onecolumn,12pt,superscriptaddress]{revtex4}

\usepackage{graphicx,color}
\usepackage{amsmath}
\usepackage{amssymb}
\begin{document}

\title{Statistical Mechanics of an Ideal Gas of Non-Abelian Anyons}

\author{Francesco Mancarella}
\affiliation{SISSA and INFN, Sezione di Trieste, via Bonomea 265, 
I-34136 Trieste, Italy}

\author{Andrea Trombettoni}
\affiliation{CNR-IOM DEMOCRITOS Simulation Center, Via Bonomea 265, I-34136 Trieste, Italy}
\affiliation{SISSA and INFN, Sezione di Trieste, via Bonomea 265, 
I-34136 Trieste, Italy}

\author{Giuseppe Mussardo}
\affiliation{SISSA and INFN, Sezione di Trieste, via Bonomea 265, 
I-34136 Trieste, Italy}
\affiliation{International Centre for Theoretical Physics (ICTP), 
Strada Costiera 11, I-34151, Trieste, Italy}

\begin{abstract}
We study the thermodynamical properties of an ideal gas of 
non-Abelian Chern-Simons particles and we compute the second virial 
coefficient, considering the effect of general 
soft-core boundary conditions for the two-body  
wavefunction at zero distance. The behaviour of the second virial 
coefficient is studied as a function of the Chern-Simons coupling, the isospin quantum number and the hard-core parameters. 
Expressions for the main thermodynamical quantities 
at the lower order of the virial expansion are also obtained: we find that at this order the relation between the internal energy 
and the pressure is the same found (exactly) for 2D Bose and Fermi ideal gases. 
A discussion of the comparison of obtained findings with available results 
in literature for systems of hard-core non-Abelian Chern-Simons particles is also supplied.
\end{abstract}

\maketitle

\begin{small}\textbf{Keywords:}\end{small} \begin{footnotesize} High energy physics;\, Condensed matter physics;\, Fractional statistics;\, Anyon thermodynamics;\, Chern-Simons theory;\, Virial expansion. \end{footnotesize}\\

\begin{small}\textbf{PACS number:}\end{small} 05.30.Pr\\

\begin{small}\textbf{Contact information:}\end{small}\\
\begin{small}Francesco Mancarella (corresponding author):\end{small} \begin{footnotesize}mancarel@sissa.it; \; Phone: +39 040 3787 273 / Fax: +39 040 3787 528 ; Postal Address: SISSA, via Bonomea 265, I-34136 Trieste, Italy ; \end{footnotesize}\\
\begin{small}Andrea Trombettoni:\end{small} \begin{footnotesize}andreatr@sissa.it ;\end{footnotesize} \\
\begin{small}Giuseppe Mussardo:\end{small} \begin{footnotesize}mussardo@sissa.it .\end{footnotesize}

\newpage

\section{Introduction} 
Unlike ordinary three-dimensional systems, quantum two-dimensional systems of indistinguishable particles allow for generalized braiding statistics. A celebrated generalization of the usual bosonic and fermionic quantum statistics is provided in two dimensions by Abelian anyons, for which a phase factor multiplying the scalar wavefunction is associated to elementary braiding operations 
\cite{Wilczek90,Khare05,Nayak08}
\begin{equation}
\psi(z_1,..,z_i,\ldots,z_j,..,z_n) \,=\,e^{i \pi\alpha} \, \psi(z_1,..,z_j,\ldots,z_i,..,z_n)\,\,\,.
\end{equation}
Anyons, first studied in \cite{Leinaas77,Goldin80,Wilczek82}, 
were later associated to the physics of the fractional quantum Hall effect  
\cite{Wilczek90}. 
Abelian anyon statistics of the simplest QH states, at filling factors $\nu=1/(2p+1)$  
were derived from a microscopic theory \cite{Arovas84}: since then, the study of the properties of Abelian anyons and the applications to the QHE have been in the following decades subject of an intense and continuing interest 
\cite{Lerda92,Forte92,Yoshioka02,Martina05,Jain07}. 

A further generalization of the bosonic and fermionic statistics is represented by non-Abelian anyons, described by a multi-component wavefunction $\psi_a(z_1,\ldots,z_n)$ ($a=1,2,\ldots,g$) which undergoes a linear unitary transformation under the effect of braiding
$\sigma_i$ which exchanges the particles at the positions $z_i$ and $z_{i+1}$  
\begin{equation}
\psi_a \rightarrow [\rho(\sigma_i)]_{ab} \, \psi_b \,\,\,,
\end{equation} 
where $\rho(\sigma_i)$ are $g \times g$ dimensional unitary matrices which do not commute among themselves, $[\rho(\sigma_i)]_{ab} [\rho(\sigma_j)]_{bc} \neq [\rho(\sigma_j)]_{ab} [\rho(\sigma_i)]_{bc}$ \cite{Nayak08}. 

Abelian and non-Abelian anyons respectively correspond to one-dimensional and higher-dimensional representations of the braid group: with respect to parastatistics, non-Abelian anyons represent the counterpart of the generalization represented by Abelian anyons with respect to ordinary Bose and Fermi statistics. Non-Abelian anyons naturally appear in the description of a variety of physical phenomena, ranging from the fractional QHE \cite{Moore91,Nayak08} to the scattering of vortices in spontaneously broken gauge theories \cite{Wilczek90-2,Bucher91,Lo93}, the (2+1)-dimensional gravity \cite{'tHooft88,Deser88,Carlip89} and the alternation and interchange of $e/4$ and $e/2$ period interference oscillations in QH heterostructures 
\cite{Willett10}. 

%The investigation of the proprieties of non-Abelian anyons is currently attracting a wide attention, both {\em per se} for the peculiarities of such strongly correlated states, but also for their relevance for the possible implementation of topological quantum computation schemes \cite{Nayak08}. 
%The reasons for such an interest are related to the fact that, using the braiding rules of non-Abelian anyons (such as the Fibonacci anyons \cite{Nayak08}), one can implement all the operations needed for universal quantum computation \cite{Nielsen00}: 
%the efficiency of schemes to perform qubit operations has been discussed in 
%\cite{Simon06,Burrello10}. 
%The main point is that non-Abelian excitations are topological, and therefore expected to be rather robust with respect 
%to the decoherence caused by the interaction with the environment. Strong evidences of the non-Abelian nature of the excitations in the $5/2$-state have been found in QH systems \cite{Willett09,An11}. 
%Furthermore, the recent realization of synthetic gauge fields 
%in systems of ultracold atoms \cite{Lin09} opens to the possibility of implementing theoretical proposals for the realization and manipulation of non-Abelian anyons in such systems 
%\cite{Aguado08,Dutta10,Roncaglia10,Burello10-2} (see more references in 
%the reviews \cite{Cooper08,Viefers08}). 
%Proposals for stabilizing non-Abelian anyons on the edges of 
%fractional quantum Hall states by proximity-coupling 
%to superconductors and/or ferromagnets \cite{Lindner12,Clarke12} have been very 
%recently discussed in literature.

The non-Abelian anyons studied in this work are non-Abelian Chern-Simons (NACS) spinless particles. The NACS particles, which are pointlike sources mutually interacting via a topological non-Abelian Aharonov-Bohm effect \cite{Dunne99},  carry non-Abelian charges and non-Abelian magnetic fluxes, so that they acquire fractional spins and obey braid statistics as non-Abelian anyons. 
More specifically, our models are described by the Hamiltonian (\ref{hamiltonianadelmodello}) which involves the isovector operators $Q_{\alpha}^a$ in a representation of isospin $l$, where $\alpha =1,2,\ldots, N$ refers to any of the $N$ particles of the system. With respect to the index $\alpha$ which labels the particles, these operators commute one to the other. Correspondingly, the quantum dimension of our anyonic systems is an integer number, contrary to what happens, for instance, in the Fibonacci anyons used to implement topological quantum computation \cite{Nayak08}, whose quantum dimension is instead an irrational number. Futhermore the NACS systems studied in this paper are gapless in the thermodynamic limit, contrary to the Fibonacci anyons or alike which have a gap in the bulk.

The study of equilibrium properties of two-dimensional anyonic systems is in general a nontrivial and highly interesting task: indeed, the anyonic statistics incorporate the effects of interaction in microscopic bosonic or fermionic systems (statistical transmutation) so that the determination of thermodynamical properties of non-interacting anyons is at least as much as difficult as the similar computation in ordinary interacting gas. This is a reason for which the investigation of equilibrium properties of a free gas of anyons called for an huge amount of efforts and work \cite{Khare05-2}, the other reason of course being that the thermodynamics of a system of free anyons is the starting point - paradigmatic for the simplicity of the model - for the understanding of the thermodynamics of more complicated interacting anyon gas.  

The two-dimensional gas of free Abelian anyons whose wavefunction fulfills hard-core wavefunction boundary conditions has been studied by Arovas, Schrieffer, Wilczek, and Zee \cite{Arovas85} in its low-density regime by taking its virial expansion. In particular, they found the exact expression for the second virial coefficient, that turns out to be 
periodic and non-analytic as a function of the statistical parameter. 
Results for higher virial coefficients of the free Abelian gas are also  
available in literature: different approaches have been used, including the semiclassical approximation \cite{Bhaduri91} and Monte Carlo computations 
\cite{Myrheim93} (for more references see \cite{Khare05,Dasnieres92}).  
Useful results can be found by perturbative expansions in powers of the 
statistical parameter $\alpha$: exact expressions for the first three terms of the expansions in powers of $\alpha$ are available for each of the first six virial coefficients 
\cite{Dasnieres92-2,Bernard92}. The second virial coefficient is the only one presenting - in each of the Bose points - cusps in the statistical parameter $\alpha$ \cite{Khare05}, i.e.  
none of the higher virial coefficients have terms at order $\alpha$ \cite{Sen91,Comtet91}. 
Furthermore, a recursive algorithm permits to compute the term in $\alpha^2 $ of all 
the cluster and virial coefficients \cite{Sen91,Comtet91,McCabe91,Emparan93}.  

The results for the virial coefficients of the free gas of Abelian anyons 
quoted in the previous paragraph are obtained considering a many-body anyonic 
wavefunction fulfilling hard-core boundary conditions, i.e. a wave function which vanishes in 
correspondence of coincident points in the configuration space of the set 
of anyons. The generalization obtained by removing such an 
hard-core constraint has been 
studied for Abelian anyons \cite{Bourdeau92,Giacconi96,Kim98} and a family of anyon models can be associated to the different boundary conditions of the same Hamiltonian. 
These models are obtained within the frame of the quantum-mechanical method of the self-adjoint extensions of the Schr\"odinger anyonic Hamiltonian. In the following we will refer to anyons without the constraint of hard-core conditions as {\em "soft-core"} or {\em "colliding"} anyons. The mathematical arguments underlying the possibility of such a generalization 
were discussed in \cite{Bourdeau92}, and the second virial coefficient of soft-core 
Abelian anyons was studied in \cite{Giacconi96,Kim98}. 
The corresponding self-adjoint extensions for the non-Abelian anyonic theory have been thoroughly discussed \cite{Bak94,Amelino95,Lee97}. 
We stress that it is not easy, in general, to extract the parameters of emerging effective 
(eventually free) anyonic models from the microscopic Hamiltonians, and then the introduction of soft-core conditions may provide useful parameters which have to be fixed via the comparison between the results of the anyonic models and 
the computations done in the underlying microscopic models.

For non-Abelian anyons, a study of the thermodynamical properties  
in the lowest Landau level of a strong magnetic field 
has been performed \cite{Polychronakos00}, showing that the 
virial coefficients are independent of the statistics. 
The theory of non-relativistic matter with non-Abelian 
Chern-Simons gauge interaction in $(2+1)$ dimensions was studied 
adopting a mean field approximation in the 
current-algebra formulation already applied to the Abelian anyons and 
finding a superfluid phase \cite{Cappelli95}. 

In comparison with the Abelian case, the thermodynamics of a system of 
free non-Abelian anyons appears to be much harder to study and 
all the available results are for hard-core boundary conditions 
\cite{Lo93-2,Lee95,Hagen96,Lee96}, with - at the best of our knowledge - 
no results (even for the second virial coefficient) for soft-core non-Abelian anyons.

The reason of this gap is at least twofold: from one side, for the difficulties, both analytical 
and numerical, in obtaining the finite temperature equation of state for non-Abelian anyons (see the discussion in \cite{Khare05-2}); from another side, because most of the efforts have been focused in the last decade on the study of two-dimensional systems which are gapped in the bulk and gapless on the edges, as for the states commonly studied for the fractional 
quantum Hall effect, while, on the contrary, the two-dimensional free gas of anyons is gapless. However, there is by now a mounting interest in the study of three-dimensional topological insulators, systems gapped in the bulk, but having protected conducting gapless states on their edge or surface \cite{Hasan10}: exotic states can occur at the surface of a three-dimensional topological insulator due to an induced energy gap, and a 
superconducting energy gap leads to a state supporting Majorana fermions, 
providing new possibilities for the realization of topological quantum computation. This surging of activity certainly calls for an investigation of the finite temperature properties of general gapless 
topological states on the two-dimensional surface of three-dimensional topological insulators and superconductors.  

In this paper we focus on the study of the thermodynamics of an ideal gas of a general class of NACS particles in presence of general soft-core boundary conditions: explicit results are found for the second virial coefficient. Results for hard-core non-Abelian anyons, which is a limiting case of soft-core conditions, are presented too. 
The article is structured as follows: in Section \ref{themodel} we introduce 
the NACS model studied in the paper and, as an introduction to the subsequent discussion,  in Section \ref{abeliananyons} we briefly recall  the results for an ideal gas of hard-core Abelian anyons and we present in detail the general soft-core version of the Abelian anyonic model. The properties of the virial expansion are also reviewed and the monotonic behaviour of the second virial coefficient $B_2$ with respect to the statistical parameter is taken in exam as the hard-core parameter changes: we observe, in particular, that for a narrow range of the soft-core parameter $B_2$ can be non-monotonic. In Section \ref{nonabeliananyons} we define the NACS model and we explicitly 
present the set of soft-core parameters associated to the most general boundary conditions of the wave-functions. In Section \ref{hardcorecase} the coefficient $B_2$ 
is evaluated for a system of NACS particles with hard-core boundary conditions: we compare
our results with previous determination of this quantity and we make 
some comments about limit cases. 
In Section \ref{generalsoftcorecase} we study $B_2$ for non-Abelian anyons 
when soft-core wavefunction boundary conditions are allowed, 
with special attention to the case of isotropic boundary conditions.  
In Section \ref{otherthermod} we summarize the virial expansions for 
the ideal gas of NACS particles endowed with general 
boundary conditions of the wave functions. Our conclusion are discussed 
in Section \ref{conclusions}. Finally the Appendices deal with some technical details 
of the main text and with the energy spectrum in the soft-core case.

\section{The Model}\label{themodel} 
In this Section we introduce 
the Abelian and non-Abelian models studied in the paper: in 
Section \ref{abeliananyons} we first briefly remind 
the well-known results for the thermodynamics of the ideal gas of hard-core Abelian anyons. The general soft-core version of the Abelian anyonic model 
is then introduced, and the behaviour of the 
second virial coefficient is studied as a function of the defined 
hard-core parameter. In Section \ref{nonabeliananyons} 
we define the NACS model, whose second virial coefficient will be derived 
and studied in the next Section.

\subsection{Abelian Anyons}\label{abeliananyons} 
The thermodynamics for a system of identical Abelian anyons has 
been developed starting with the seminal paper \cite{Arovas85}, 
in which the exact quantum expression for the second virial 
coefficient is derived: 
\begin{equation} 
B_2^{h.c.}(2j+\delta,T)=-\frac{1}{4}\lambda_T^2+\vert\delta\vert\lambda_T^2-\frac{1}{2} \delta^2\lambda_T^2\,\,\,.\label{BdiArovas} 
\end{equation} 
Eq. (\ref{BdiArovas}) holds for an ideal gas of anyons whose wavefunction fulfills 
hard-core wavefunction boundary conditions. In (\ref{BdiArovas}) 
$\alpha=2j+\delta$, where $\alpha$ 
represents the statistical parameter of the anyons \cite{Khare05}, 
$j$ is an integer 
and $\vert\delta\vert\leq 1$. We remind that $\alpha=0$ and $\alpha=1$ 
corresponds respectively to free two-dimensional spinless bosons and fermions 
\cite{Khare05}. Furthermore $\lambda_T$ is the thermal wavelength defined as
\begin{equation} 
\lambda_T=\left(\frac{2 \pi \hbar^2}{M k_B T} \right)^{1/2}\,\,\,.\label{TDBBWL} 
\end{equation} 
As discussed in statistical mechanics textbooks, the virial expansion is done in powers of $\rho \lambda_T^2$ (where $\rho$ is the density and $M$ is the mass of the particles) 
and in the low-density, high-temperature regime, the second virial coefficient 
gives the leading contribution to the deviation of the equation of state 
from the non-interacting case, as a result of rewriting the grand canonical partition function as a cluster expansion \cite{Mayer77,Huang87}.

The virial coefficient (\ref{BdiArovas}) turns out to be a simple, periodic (with period $2$) 
but non-analytic function of the statistical parameter $\alpha$, showing cusps in correspondence of all its bosonic points. This quantity has been evaluated by different methods: one of them consists in an hard-disk-type regularization of the two-anyonic spectrum while another one is a based on path-integral approach yielding 
the two-body partition function, carried on by identifying the Lagrangian of the system with the one relative to the Bohm-Aharonov effect \cite{Aharonov59}. Eq. (\ref{BdiArovas}) is also retrieved by heat kernel methods, i.e. discretizing the two-particle spectrum through the introduction of a harmonic regulator potential and then directly considering the problem in 
the continuum \cite{Comtet89}. Finally, another method to get Eq. (\ref{BdiArovas}) is to 
use a semiclassical method, which nevertheless produces the exact quantum result \cite{Bhaduri91,Khare05-2}. Exact results for higher virial coefficients are not known, but a fair amount of information is available both for the third virial coefficient and for 
higher virial coefficients \cite{Khare05}.

The expression (\ref{BdiArovas}) is the exact quantum result for the hard-core case, corresponding to impose the vanishing of the two-anyonic wavefunctions in the coincident points (the limit configurations for which the coordinates of two anyons coincide). 
However, any arbitrary boundary condition for the wave-function is in 
principle admissible: in general it is the comparison with results from the microscopic 
interacting Hamiltonian that should fix the relevant boundary conditions 
to be imposed. The second virial coefficient for Abelian anyons in this 
general case has been studied in \cite{Giacconi96,Moroz96,Kim98}.

By relaxing the regularity requirements, allowing wave-functions to diverge for vanishing relative distance $r$ between the anyons according to the method of self-adjoint extensions, it is possible to obtain a one-parameter family of boundary conditions. The hard-core limit corresponds to scale-invariance in a field theoretical approach \cite{Hagen85,Jackiw90,Bourdeau92,Kim95,Kim97}, where the scale can be precisely related to the hard-core parameter that will be defined below. The study of $B_2$ shows that the 
results for hard-core case are rather peculiar: for instance, the cusps at the bosonic points of the hard-core case are a special feature of the scale-invariant limit, which is however 
absent for all the soft-core cases. On the contrary, cusps are generated at all the fermionic points for all the wavefunction boundary conditions, except just for the hard-core case. 

The {\em relative} two-body Hamiltonian for a free system of anyons with statistical 
parameter $\alpha$, 
written in the bosonic description, is of the form \cite{Khare05}  
\begin{equation} 
H_{rel}=\frac{1}{M}(\vec{p}-\alpha \vec{A})^2\,\,\,, 
\end{equation} 
where $\vec{A}=(A^1,A^2)$ and $A^i\equiv\frac{\epsilon^{ij}x^j}{r^2}$ 
($i=1,2$ and $\epsilon^{ij}$ is the completely antisymmetric tensor). The corresponding single-particle 
partition function of the relative dynamics is $Z_{rel}={\rm Tr } e^{-\beta H_{rel}}$, where $\beta=1/k_B T$. 
%The second virial coefficient is   
%\begin{equation}   
%B(T) = A \left( \frac12 - \frac{Z_2}/{Z_1^2} \right)\,\,\,, 
%\end{equation}
%with $Z_1 =A/\lambda_T^{2}$ and $Z_2= 2  A Z_{rel} / \lambda_T^{2}$. 
In order to proceed with the choice of a given self-adjoint extension,  
one has to define the space over which the trace above is performed. 
If we consider the radial component $R_n$ of the relative wave-function 
$\psi$, 
the Schr\"odinger equation takes the form  
\begin{equation} \label{eqradiale} 
\frac{1}{M}\left[-\frac1r \frac{d}{dr}r\frac{d}{dr} + \frac{(n+\alpha)^2}{r^2}\right] 
R_n(r) = ER_n(r) \equiv \frac{k^2}{M}R_n(r)\,\,\,, 
\end{equation} 
with $n$ even (choosing the bosonic description) \cite{Khare05}. Without any loss of generality, 
the statistical parameter can be chosen as $\alpha \in [-1,1]$. Eq. (\ref{eqradiale}) is the Bessel equation and its general solution is given in terms of the Bessel functions: 
\begin{equation} \label{solution} 
R_n(r) = A J_{|n+\alpha|}(kr) + B J_{-|n+\alpha|}(kr)\,\,\,.
\end{equation} 
For $n\neq 0$ the constant $B$ must vanish in order to satisfy 
the normalization of the relative wave-function, while for $n=0$ ($s$-wave) arbitrary constants $A$, $B$ are allowed. This yields a one-parameter family of boundary conditions for the $s$-wave solution: 
\begin{equation} \label{sol} 
R_0(r) = (const.) \, \left[ 
        J_{|\alpha|}(kr) + \sigma \left(\frac{k}{\kappa}\right)^{2|\alpha|}J_{-|\alpha|}(kr) 
        \right]\,\,\,, 
\end{equation}
where $\sigma=\pm1$ and $\kappa$ 
is a scale introduced by the boundary condition. 

We will refer to 
\begin{equation}\label{hard-core-par}
\varepsilon \equiv \frac{\beta \kappa^2}{M}
\end{equation} 
as the {\em hard-core parameter} of the gas. 
For $\varepsilon\rightarrow \infty$ with $\sigma=+1$ we retrieve the hard-core case ($\psi(0)=0$). If $\sigma=-1,$ in addition to the solution (\ref{sol}), there is a bound state with energy $E_B=-\varepsilon k_B T=-\kappa^2 / M$ and wavefunction  
\begin{equation} 
R_0(r)= (const.) \, K_{|\alpha|}(\kappa r)\,\,\,,\label{boundstate} 
\end{equation} 
being $K_\alpha(x)$ the modified Bessel function of the second type. 
By proceeding as in \cite{Arovas85}, and observing that only the $s$-wave 
energy spectrum is modified with respect to the hard-core case, 
one gets that the second virial coefficient for a generic soft-core is 
given by  
\begin{equation} \label{virial} 
B_{2}^{s.c.}(T)=B_2^{h.c.}(T) - 2\lambda_T^2\left\{ e^{-\beta E_B}\theta(-\sigma) +  
    \lim_{R\rightarrow\infty}\sum_{s=0}^\infty \left[  
        e^{-\beta\frac{\tilde{k}_s^2}{M}}-e^{-\beta\frac{k_{0,s}^2}{M}}\right]\right\}\,\,\,, 
\end{equation}
where $\theta(x)$ is the Heaviside step function, 
$k_{0,s}R$ is the $s$-th zero of $J_{\vert\alpha\vert}(kR)=0$, 
$\tilde{k}_s R$ is the $s$-th zero of (\ref{sol}), and $B_2^{h.c.}$ 
is the hard-core result (\ref{BdiArovas}).  
It is possible to rewrite Eq. (\ref{virial}) in an integral form \cite{Kim98} 
as  
\begin{equation} 
\label{explicitintegralform} 
B_2^{s.c.}(T)=B_2^{h.c.}(T)- 2 \lambda_T^2 
\left\{ 
e^\varepsilon \theta(-\sigma)  
+\frac{\alpha\sigma}{\pi} \sin{\pi\alpha} 
\int_0^\infty \frac{dt e^{-\varepsilon t} t^{|\alpha|-1}}{1+2\sigma\cos{\pi\alpha}\;t^{|\alpha|}+t^{2|\alpha|}} \right\}\,\,\,.
\end{equation} 
For $\varepsilon\gg 1$ one gets
\begin{equation} 
B_{2}^{s.c.}(T)=B_2^{h.c.}(T)- 2 \lambda_T^2 \;[e^\varepsilon \theta(-\sigma)+\frac{\sigma}{\pi}\frac{\Gamma(\vert\alpha\vert+1)}{\varepsilon^{|\alpha|}}\sin{\pi|\alpha|}+\cdots ] 
\end{equation} 
while for $\varepsilon\ll 1$  
\begin{equation} 
B_{2}^{s.c.}(T)=B_2^{h.c.}(T)- 2 \lambda_T^2 |\alpha|\;(1-\sigma \, \varepsilon^{|\alpha|}+\cdots)\,\,\,.   
\end{equation} 
Near the bosonic point $\alpha=0$ one has for $\varepsilon\neq 0$: 
\begin{equation} 
B_{2}^{s.c.}(T)=-\left[\frac{1}{4}+2 \nu(\varepsilon)\theta(-\sigma)\right]\,\lambda_T^2+O(\alpha^2)\,\,\,, 
\nonumber 
\end{equation} 
where $\nu(\varepsilon)$ is the Neumann function defined by 
$$ 
\nu(\varepsilon)=\int_0^\infty \frac{dt\,\varepsilon^t}{\Gamma(t+1)}\,\,\,. 
$$
From this expression one sees that $B_2^{s.c.}(T)$ is a smooth function of the statistical parameter near $\alpha=0$. On the contrary, near the fermionic point $|\alpha|=1$ one has  
\begin{equation} 
B_{2}^{s.c.}(T)=\left[-\frac{1}{2}(1-\vert\alpha\vert)^2+\frac{1}{4}-2 e^{-\sigma \varepsilon}\right]\,\lambda_T^2+f_{\sigma}(\varepsilon)(1-|\alpha|) +\cdots\,\,\,, 
\nonumber 
\end{equation} 
where $f_{\sigma}(\varepsilon)$ ($\sigma=\pm1$) are functions of 
$\varepsilon$ [not reported here], so that in general the soft-core case presents a cusp at 
$|\alpha|=1.$  

\begin{figure}[t]
\centerline{\scalebox{0.35}{\includegraphics{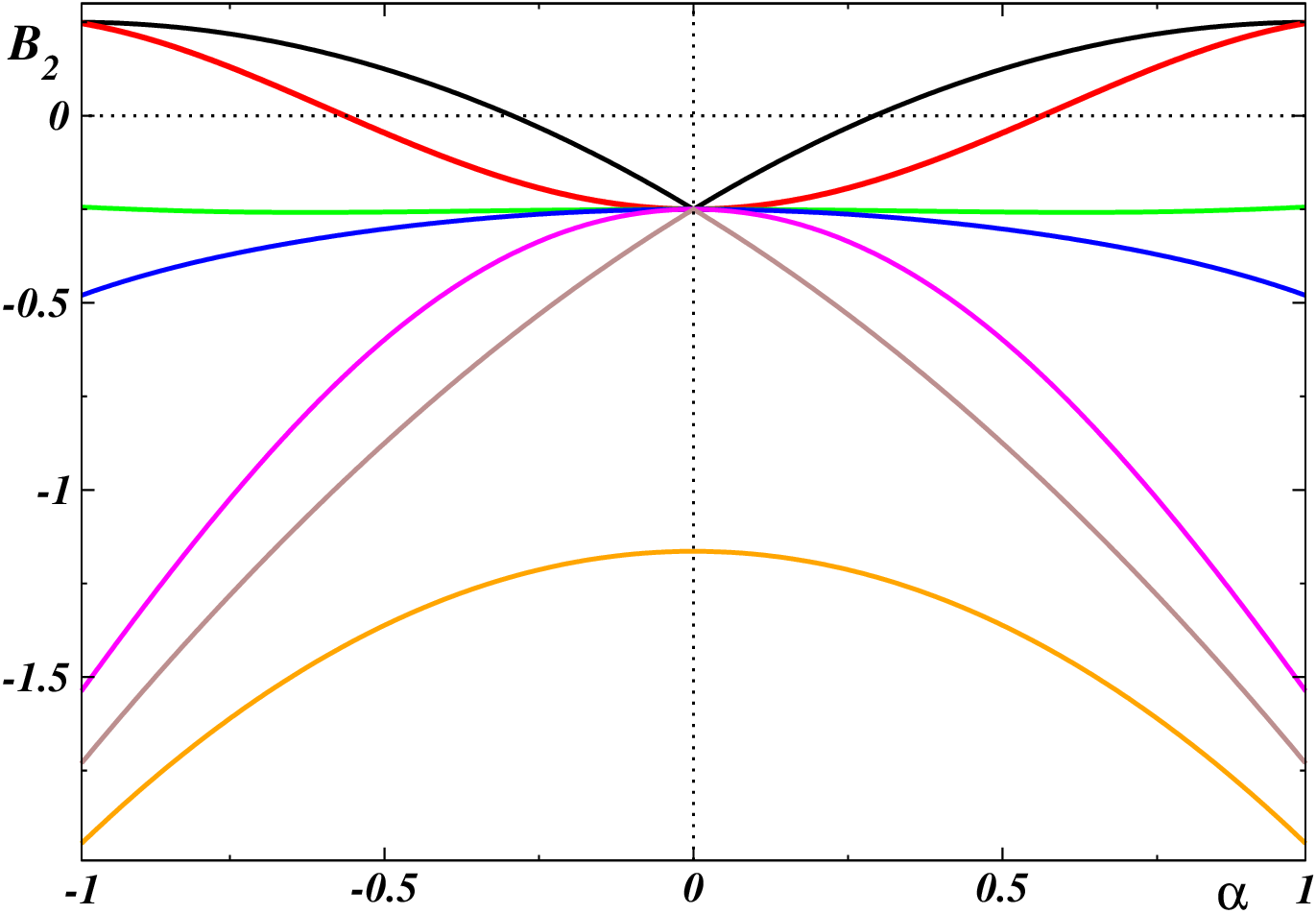}}}
%\vspace{0.5cm}
\caption{$B_2$ (in units of $\lambda_T^2$) 
vs. the statistical parameter $\alpha$ for different values 
of the hard-core parameter $\varepsilon$ for Abelian anyons. The $6$ 
upper curves are obtained for $\sigma=1$: from top to the bottom 
$\varepsilon$ takes the values 
$\infty$ (hard-core), $10$, $1.4$, $1$, $0.1$ and $0$. 
The curve below all those is obtained for $\sigma=-1$ 
and $\varepsilon=0.1$ and it has a shifted value at the bosonic points 
(the dotted lines just denote the $x$ and $y$ axes). 
We remind that $B_2$ is periodic in $\alpha$ with period $2$, so we plot 
it in the interval $[-1,1]$. $B_2$ is symmetric with respect to all the 
integer value of $\alpha$: we see from the figure 
the difference in the position of the cusps of the patterns, 
between the hard-core and the soft-core cases.}
\label{fig1}
\end{figure}

\begin{figure}[t]
\centerline{\scalebox{0.35}{\includegraphics{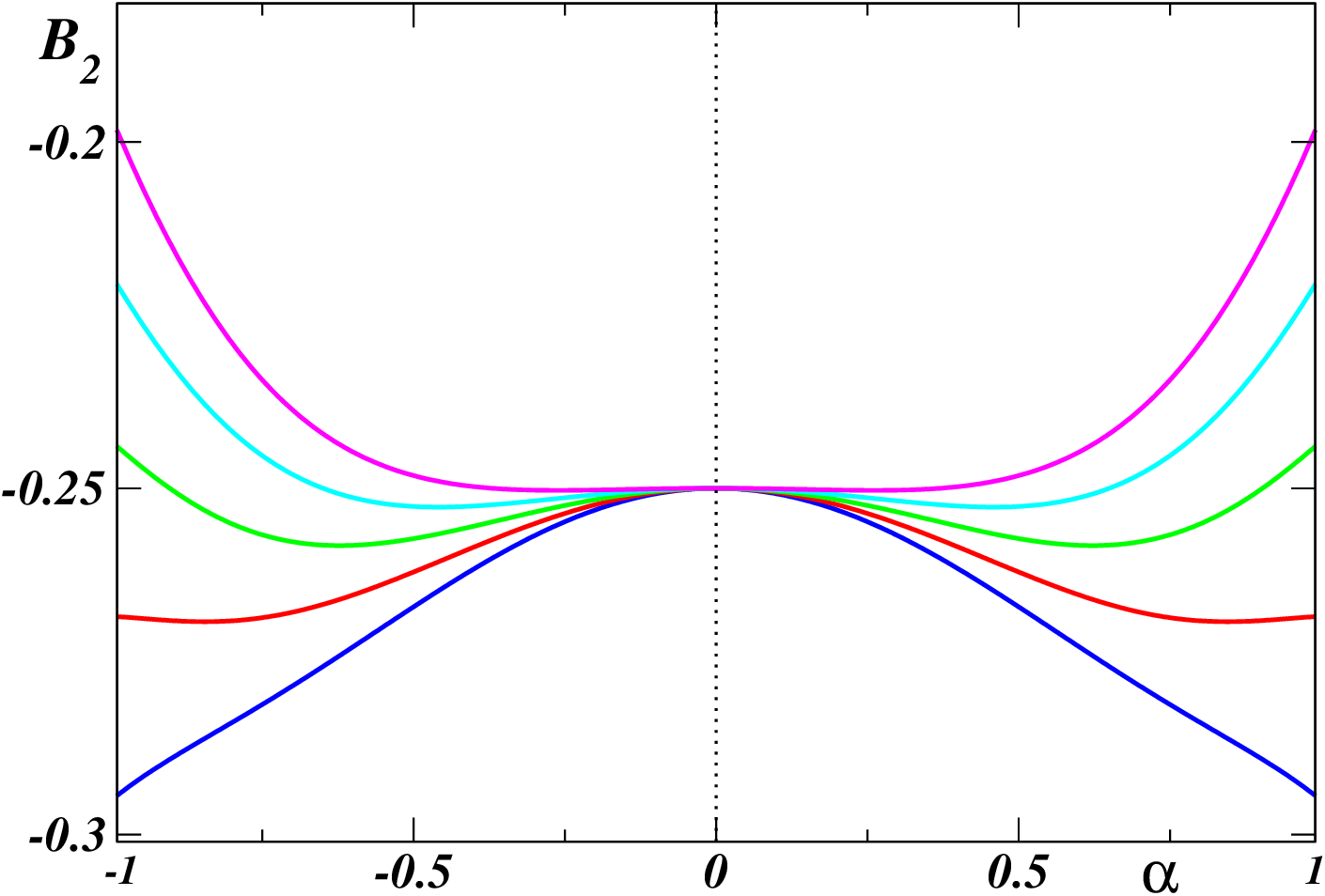}}}
%\vspace{0.5cm}
\caption{The narrow range of values of the hard-core parameter 
$\varepsilon$ within which the second virial coefficient $B_2$ 
(for Abelian anyons)  
is a non-monotonous function of $\alpha$ in the interval $[0,1]$. 
The parameters are $\sigma=+1$ for all the curves and 
$\varepsilon=1.50, \,1.45,\, 1.40,\, 1.35,\, 1.30 $ from top to bottom 
($B_2$ is expressed in units of $\lambda_T^2$).}
\label{fig2}
\end{figure}

The virial coefficient $B_2^{s.c.}$ is plotted in Fig.\ref{fig1} 
for some values of the hard-core parameter $\varepsilon$. 
The plot of the second virial coefficient clearly exhibits 
its smoothing in the bosonic points and its sharpening in the fermionic ones, 
as soon as the hard-core condition is relaxed. We observe that 
the restriction of $B_2^{s.c.}$ over the interval $[0,1]$ 
is not a monotonic function of $\alpha$ for each $\varepsilon$. 
This not-monotonicity happens around $\varepsilon\sim 1.4$ 
in a narrow range of values of $\varepsilon$ ($1.344<\varepsilon<1.526$), 
as emphasized in Fig.\ref{fig2}.

\subsection{Non-Abelian Anyons}\label{nonabeliananyons} 
The main part of the present paper deals with the study of the 
low-density statistical mechanics properties of a two-dimensional gas of 
$SU(2)$ NACS spinless particles. 
The NACS particles are pointlike sources mutually interacting 
via a topological non-Abelian Aharonov-Bohm effect \cite{Dunne99}. 
These particles carry non-Abelian charges and non-Abelian magnetic fluxes, so that they acquire fractional spins and obey braid statistics as non-Abelian anyons. 

In order to proceed with the computation of the second virial coefficient 
of a free gas of NACS particles, we first introduce the NACS quantum mechanics 
\cite{Guadagnini90,Verlinde91,Lee93,Kim94,Bak94} considering 
the general frame of soft-core NACS particles \cite{Amelino95,Lee97}. 
The Hamiltonian describing the dynamics of the $N$-body system of free 
NACS particles can be derived by a Lagrangian with a Chern-Simons term and 
a matter field coupled with the Chern-Simons gauge term \cite{Bak94}: the 
resulting Hamiltonian reads
\begin{equation} 
{H}_N=-\sum_{\alpha=1}^{N} \frac{1}{M_\alpha}\left(\nabla_{\bar 
z_\alpha}\nabla_{z_\alpha}  +\nabla_{z_\alpha}\nabla_{\bar 
z_\alpha}\right)
\label{hamiltonianadelmodello}
\end{equation}
where $M_\alpha$ is the mass of the $\alpha$-th particles, 
$\nabla_{\bar z_\alpha}=\frac{\partial}{\partial \bar z_\alpha}$ and 
$$
\nabla_{z_\alpha}=\frac{\partial}{\partial z_\alpha}  +\frac{1}{2\pi 
\kappa} \sum_{\beta\not=\alpha} \hat Q^a_\alpha \hat Q^a_\beta \frac{1}{ 
z_\alpha -z_\beta}\,\,\,.
$$
In Eq. (\ref{hamiltonianadelmodello}), $\alpha = 1, \dots, N$ 
labels the particles, $(x_\alpha, y_\alpha)=(z_\alpha+\bar z_\alpha, 
-i(z_\alpha-\bar z_\alpha))/2$ are their spatial coordinates, 
and $\hat Q^a$'s are the isovector operators in a representation of isospin $l$.  The quantum number 
$l$ labels the irreducible representations of the group of the rotations induced by the 
coupling of the NACS particle matter field with the non-Abelian gauge field: as a consequence, the values of $l$ are of course quantized and 
vary over all the integer and the half-integer numbers, with $l=1/2$ being  
the smaller possible non-trivial value ($l=0$ corresponds to a system of free bosons). As usual, a basis of 
isospin eigenstates can be labeled by $l$ and the magnetic quantum number $m$ (varying in the range $-l,-l+1,\cdots,l-1,l$).

The virial coefficients then depend in general on the value of the isospin quantum number 
$l$ and on the coupling $\kappa$ (and of course on the temperature $T$). 
The quantity $\kappa$ in (\ref{hamiltonianadelmodello}) 
is a parameter of the theory. In order to enforce the gauge covariance 
of the theory the condition $4 \pi \kappa \, = \, {\rm integer}$ 
has to be satisfied. In the following we denote for simplicity the 
integer $4 \pi \kappa$ by $k$:
\begin{equation}
4 \pi \kappa \, \equiv k\,\,\,.
\label{integer_kappa}
\end{equation} 
The physical meaning of $\kappa$ in the NACS model can be understood 
removing the interaction terms in $H_N$ by a similarity transformation:  
\begin{eqnarray} 
{H}_N&\longrightarrow & UH_N U^{-1}= H^{\rm free}_N = -\sum^N_\alpha 
\frac{2}{M_\alpha} \partial_{\bar z_\alpha}\partial_{z_\alpha}\nonumber\\ 
\Psi_H &\longrightarrow & U \Psi_H = \Psi_A \label{simil} 
\end{eqnarray} 
where $U(z_1,\dots,z_N)$ satisfies the Knizhnik-Zamolodchikov (KZ) 
equation \cite{kz} 
\begin{equation} 
\left(\frac{\partial}{\partial z_\alpha}  - \frac{1}{ 2\pi 
\kappa} \sum_{\beta\not=\alpha} \hat Q^a_\alpha \hat Q^a_\beta \frac{1}{z_\alpha -z_\beta}\right) U(z_1,\dots,z_N) =0\,\,\,, 
\end{equation} 
and $\Psi_H(z_1,\dots,z_N)$ stands for the wavefunction of the $N$-body system of the NACS particles in the holomorphic gauge. A comparison between the last equation and the KZ equation satisfied by the Green's function in the conformal field theory shows that $(4 \pi \kappa-2)$ corresponds to the level of the underlying $SU(2)$ current algebra. In \cite{Lee93} it is shown how $\Psi_A(z_1,\dots,z_N)$ obeys the braid statistics due to the transformation function $U(z_1,\dots,z_N),$ while $\Psi_H(z_1,\dots,z_N)$ satisfies ordinary statistics: 
$\Psi_A(z_1,\dots,z_N)$ can be then referred to as 
the NACS particle wavefunction in the anyon gauge.  

The statistical mechanics of the NACS particles can be studied 
by introducing the grand partition function $\Xi$, defined in terms of the 
$N-$body Hamiltonian $H_N$ and the fugacity $\nu$ as 
\begin{equation} 
\Xi = \sum_{N=0}^\infty \nu^N\, {\rm Tr}\, e^{-\beta H_N} \,\,\,.
\label{par1}
\end{equation} 
In the low-density regime, a cluster expansion can be applied to $\Xi$ 
\cite{Mayer77,Huang87}: 
\begin{equation} 
\Xi = \exp \left( V\sum_{n=1}^\infty b_n \nu^n\right)\,\,\,,\label{par2} 
\end{equation} 
where $V$ is the volume of the gas (of course, for 
a two-dimensional gas $V$ equals the area $A$) 
and $b_n$ is the $n$-th cluster integral, with  
\begin{equation} 
b_1 = \frac{1}{A} Z_1,\quad b_2 = \frac{1}{A}\left(Z_2 - \frac{Z_1^2}{2}\right)\,\,\, 
\end{equation} 
and $Z_N={\rm Tr}\, e^{-\beta H_N}$ being the $N$-particle partition function.

The virial expansion (i.e. the pressure expressed in powers of the density $\rho=\frac{N}{A}$) is given as  
\begin{equation} 
P= \rho k_B T\left[ 1+ B_2(T) \rho+ B_3(T) \rho^2 +\dots \right] \,\,\,, 
\end{equation} 
where $B_n(T)$ is the $n$-th virial coefficient. The second virial coefficient $B_2(T)$ is written as  
\begin{equation} 
B_2(T) \,=\, -\frac{b_2}{b_1^2} 
\,=\, A\,\left(\frac{1}{2}-\frac{Z_2}{Z_1^2}\right)\,\,\,. 
\end{equation} 
We assume that the NACS particles belong to the same isospin multiplet 
$\{|l,m>\}$ with $m=-l, \dots, l$. The quantity $Z_1={\rm Tr}\, e^{-\beta H_1}$ 
is then given by
\begin{equation} 
Z_1 = (2l+1) A/ \lambda^{2}_T \label{zone}\,\,\,. 
\end{equation}

The computation of $Z_2= {\rm Tr}\, e^{-\beta H_2}$ is discussed in 
\cite{Lee95}, where the results for the hard-core case are presented. It is convenient 
to separate the center-of-mass and relative coordinates: defining 
$Z = (z_1+z_2)/2$ and $z = z_1 -z_2$ one can write 
\begin{equation} 
H_2 = H_{\rm cm} + H_{\rm rel}=-\frac{1}{ 2\mu} \partial_Z \partial_{\bar Z} 
-\frac{1}{\mu}(\nabla_z\nabla_{\bar z} +\nabla_{\bar z}\nabla_z)\,\,\,, 
\end{equation} 
where $\mu\equiv M/2$ is the two-body reduced mass, $\nabla_{\bar z} = \partial_{\bar z}$ and 
$$
\nabla_z = \partial_z +\frac{\Omega}{z}\,\,\,. 
$$
$\Omega$ is a block-diagonal matrix given by  
$$
\Omega = \hat Q^a_1\hat Q^a_2 / (2\pi\kappa)=
\sum_{j=0}^{2l} \omega_j\otimes{I}_j\,\,\,, 
$$ 
with $\omega_j\equiv\frac{1}{4\pi\kappa} 
\left[j(j+1)-2l(l+1)\right]$. \; $Z_2$ can be then written as 
\begin{equation} 
Z_2= 2A  \lambda^{-2}_T Z_2^\prime \,\,\,, 
\label{hamrel:a}
\end{equation}
where $Z_2^\prime ={\rm Tr}_{\rm rel}\, e^{-\beta H_{\rm rel}}$. 
The similarity transformation 
$G(z,\bar z)  = \exp\left\{-\frac{\Omega}{2}\ln(z\bar z)\right\}$,  
acting as
\begin{eqnarray} 
H_{\rm rel} &\longrightarrow & 
H_{\rm rel}^\prime = G^{-1} H_{\rm 
rel} G,\nonumber\\ 
\Psi(z,\bar z) &\longrightarrow & 
\Psi^\prime(z,\bar z) =G^{-1} \Psi(z,\bar z)\,\,\,, \label{sim} 
\end{eqnarray} 
gives rise to an Hamiltonian $H_{\rm rel}^\prime$ 
manifestly Hermitian  
%\begin{eqnarray} 
%H_{\rm rel}^\prime &=& -\frac{1}{\mu}(\nabla_z^\prime\nabla_{\bar z}^\prime 
%+ \nabla_{\bar z}^\prime\nabla_z^\prime),\label{trham}\\ 
%\nabla_z^\prime &=& \partial_z + \frac{\Omega}{2}\frac{1}{z},\quad 
%\nabla_{\bar z}^\prime=\partial_{\bar z} -\frac{\Omega}{2}\frac{1}{\bar z}. 
%\nonumber 
%\end{eqnarray}  
and leaves invariant $Z^\prime_2$.
% = {\rm Tr}_{\rm rel}\, e^{-\beta H_{\rm rel}^\prime}.$ 
The explicit expression for $H_{\rm rel}^\prime$ is
\begin{equation} 
H_{\rm rel}^\prime = -\frac{1}{\mu}(\nabla_z^\prime\nabla_{\bar z}^\prime 
+ \nabla_{\bar z}^\prime\nabla_z^\prime)\,\,\,,\label{trham}
\end{equation} 
where $\nabla_z^\prime = \partial_z + \Omega/2z$ and 
$\nabla_{\bar z}^\prime=\partial_{\bar z} -\Omega/2 {\bar z}$. 

By rewriting $H_{\rm rel}^\prime$ in polar coordinates and projecting it onto the subspace of total isospin $j,$ its correspondence with the Hamiltonian for (Abelian) anyons in the Coulomb gauge, having statistical parameter given by $\alpha_s= \omega_j$, becomes evident: 
\begin{equation} 
H_j^\prime = -\frac{1}{2\mu}\left[\frac{\partial^2}{\partial r^2}+ 
\frac{1}{r}\frac{\partial}{\partial r}+\frac{1}{r^2}\left(\frac{\partial} 
{\partial \theta}+i\omega_j\right)^2\right]\,\,\,. 
\end{equation} 
The same analysis discussed in Section \ref{abeliananyons} shows 
that the radial factor of the $j,j_z-$ component of the relative 
$(2l+1)^2-$vector wavefunction $\psi=e^{i n\theta} R_n(r)$ 
obeys the Bessel equation 
\begin{equation} 
\frac{1}{M}\left[-\frac1r \frac{d}{dr}r\frac{d}{dr} + \frac{(n+\omega_j)^2}{r^2}\right] 
R^{j,j_z}_n(r) = ER^{j,j_z}_n(r) \equiv \frac{k^2}{M}R^{j,j_z}_n(r)\,\,\,,
\end{equation}   
whose general solution is  
\begin{equation} 
R^{j,j_z}_n(r) = A^{j,j_z}J_{|n+\omega_j|}(kr) + B^{j,j_z}J_{-|n+\omega_j|}(kr)\,\,\,. 
\end{equation} 
As already discussed in the previous Section \ref{abeliananyons}, 
$B^{j,j_z}$ can be nonzero only in the case $n=0$ ($s-$wave). Then 
the $s-$wave gives rise to a one-parameter family of boundary conditions  
\begin{equation} 
\label{nonabelianradialsoftcore} 
R^{j,j_z}_0(r) = \mbox{(const.)}\left[ 
        J_{|\omega_j|}(kr) + \sigma \left(\frac{k}{\kappa_{j,j_z}}\right)^{2|\omega_j|}J_{-|\omega_j|}(kr)\,\,\,, 
        \right]\,, 
\end{equation} 
where $\sigma=\pm 1$, and $\kappa_{j,j_z}$ is a momentum scale 
introduced by the boundary condition.  

We refer to the $(2l+1)^2$ quantities 
\begin{equation}\label{hard-core-par-NA}
\varepsilon_{j,j_z}\equiv \frac{\beta\kappa^2_{j,j_z}}{M}
\end{equation} 
as {\em hard-core parameters} of the system. 
The hard-core limit corresponds to 
$\varepsilon_{j,j_z}\rightarrow\infty$ for all $j,j_z$.  

We conclude this Section by observing that, according 
to the regularization used in \cite{Arovas85,Comtet89}, 
the second virial coefficient is defined as  
\begin{equation} 
B_2(\kappa, l, T) - B_2^{(n.i.)}(l,T) = -\frac{2\lambda_T^2}{(2l+1)^2} 
\left[Z^\prime_2 (\kappa, l, T) - Z^{\prime(n.i.)}_2 (l,T)\right]\,\,\,,\label{legamepartizioneviriale} 
\end{equation} 
where $B_2^{(n.i.)}(l,T)$ is the virial coefficient for the system with particle isospin $l$ and without interaction ($\kappa \rightarrow \infty$), 
which will be expressed in terms of the virial coefficients 
$B_2^B(T)$, $B_2^F(T)$ of the free Bose and Fermi systems with the considered 
general wavefunction boundary conditions. Furthermore, 
$Z^{\prime}_2 (\kappa, l, T) - Z^{\prime(n.i.)}_2 (l,T)$ 
is the (convergent) variation of the divergent partition function 
for the two-body relative Hamiltonian, between the interacting case in exam 
and the non-interacting limit ($\kappa \rightarrow \infty$).

\section{Second Virial Coefficient}
In this Section we present our results for the second virial coefficient of 
a free gas of NACS particles: we will first study the hard-core case 
in Section \ref{hardcorecase}, comparing in detail 
our findings with results available 
in literature \cite{Lo93-2,Lee95,Hagen96,Lee96}. We then study 
$B_2$ for non-Abelian anyons when general soft-core wavefunction 
boundary conditions, focusing the attention in particular to the isotropic boundary conditions.

\subsection{Hard-Core Case}\label{hardcorecase} 
The hard-core case is obtained in the limit 
$\varepsilon_{j,j_z}\rightarrow\infty$ for all $j,j_z$. The second virial 
coefficient has been discussed in literature, and different results 
for $B_2$ have been presented \cite{Lo93-2,Lee95,Hagen96}: the differences 
between such results have been discussed, see in particular 
Ref. \cite{Lee95} and the comment \cite{Hagen96}. Our findings 
differ 
from results presented in \cite{Lo93-2,Lee95,Hagen96}: 
in this Section, as well as in Appendices 
\ref{appendix_lo}-\ref{hagen_appendix}, a detailed  
comparison with such available results will be presented.

For hard-core boundary conditions of the relative two-anyonic 
vectorial wavefunction, the quantity $B_2^{(n.i.)}$ 
entering Eq. (\ref{legamepartizioneviriale}) is found to be 
\cite{Lee95} 
\begin{eqnarray} 
B_2^{(n.i.)}(l,T)  = \frac{1}{(2l+1)^2} \sum^{2l}_{j=0} (2j+1) 
\left[\frac{1+(-1)^{j+2l}}{2}B_2^B(T)+ \frac{1-(-1)^{j+2l}}{2}B_2^F(T)\right]\,\,\,. 
\label{deltabidue}
\end{eqnarray}  
Using for the hard-core case the known values 
$B_2^B(T)=-B_2^F(T)=-\frac{1}{4} \lambda_T^2$ \cite{Khare05-2},  
one then obtains 
\begin{eqnarray} 
B_2^{(n.i.)}(l,T)  = - \, \frac{\lambda^2_T}{4} \,  \frac{1}{2l+1}\,\,\,. 
\label{limitenoninteragente} 
\end{eqnarray}  
To proceed further, we introduce a regularizing harmonic potential 
${\cal V}=\frac{\mu}{2} \epsilon^2 r^2$ \cite{Arovas85,Khare05-2},  
whose effect is to make discrete the spectrum of $H_{j}^\prime$.  
Using the notations of \cite{Khare05}, p.48, the spectrum consists 
of the following two classes: 
$E^I_n = \epsilon (2n+1+\gamma_j)$ with degeneracy $(n+1)$, 
and $E^{II}_n = \epsilon (2n+1-\gamma_j)$ with  
degeneracy $n$, 
where $n$ is a non-negative integer and 
$\gamma_j\equiv\omega_j \, mod \, 2$. 
It follows that the regularized partition function reads
$$Z^\prime_2 (\kappa, l, T) - Z^{\prime(n.i.)}_2 (l,T)=\sum_{j=0}^{2l}(2j+1)\lim_{\epsilon\rightarrow 
0}\Biggl\{\frac{1+(-1)^{j+2l}}{2}\left\{Z_\epsilon^\prime(\gamma_j) -
Z_\epsilon^\prime(0)\right\} 
+$$ 
\begin{eqnarray}  
+\frac{1-(-1)^{j+2l}}{2} \left\{Z_\epsilon^\prime [(\gamma_j+1) \, mod \, 2] 
-Z_\epsilon^\prime(1)\right\}\Biggr\}\,\,\,, 
\end{eqnarray} 
with 
\begin{eqnarray} 
Z_\epsilon^\prime(\gamma_j)=  \sum_{n=0}^{\infty}\left[(n+1)\, e^{-\beta\epsilon(2n+1 
+\gamma_j)}+n\, e^{-\beta\epsilon(2n+1-\gamma_j)}\right]   
=\frac{1}{2} \frac{\cosh\left[\beta\epsilon(\gamma_j-1)\right]}{\sinh^2 
\beta\epsilon}\,\,\,.
\nonumber 
\end{eqnarray} 
The final result for the NACS gas in the hard-core limit 
is then the following: 
$$
B_2^{h.c.}(\kappa, l, T) =  -\frac{\lambda_T^2}{4} \frac{1}{2l+1} + 
$$
\begin{equation}\label{coeffviriale}
-\frac{\lambda_T^2}{2(2l+1)^2} \sum_{j=0}^{2l}(2j+1)\Biggl[\frac{1+(-1)^{j+2l}}{2} (\gamma_j^2-2 \gamma_j)+\frac{1-(-1)^{j+2l}}{2}[(\gamma_j+1) \, mod \, 2-1]^2\Biggr]
\,\,\,.
\end{equation} 

\begin{figure}[t]
\centerline{\scalebox{0.40}{\includegraphics{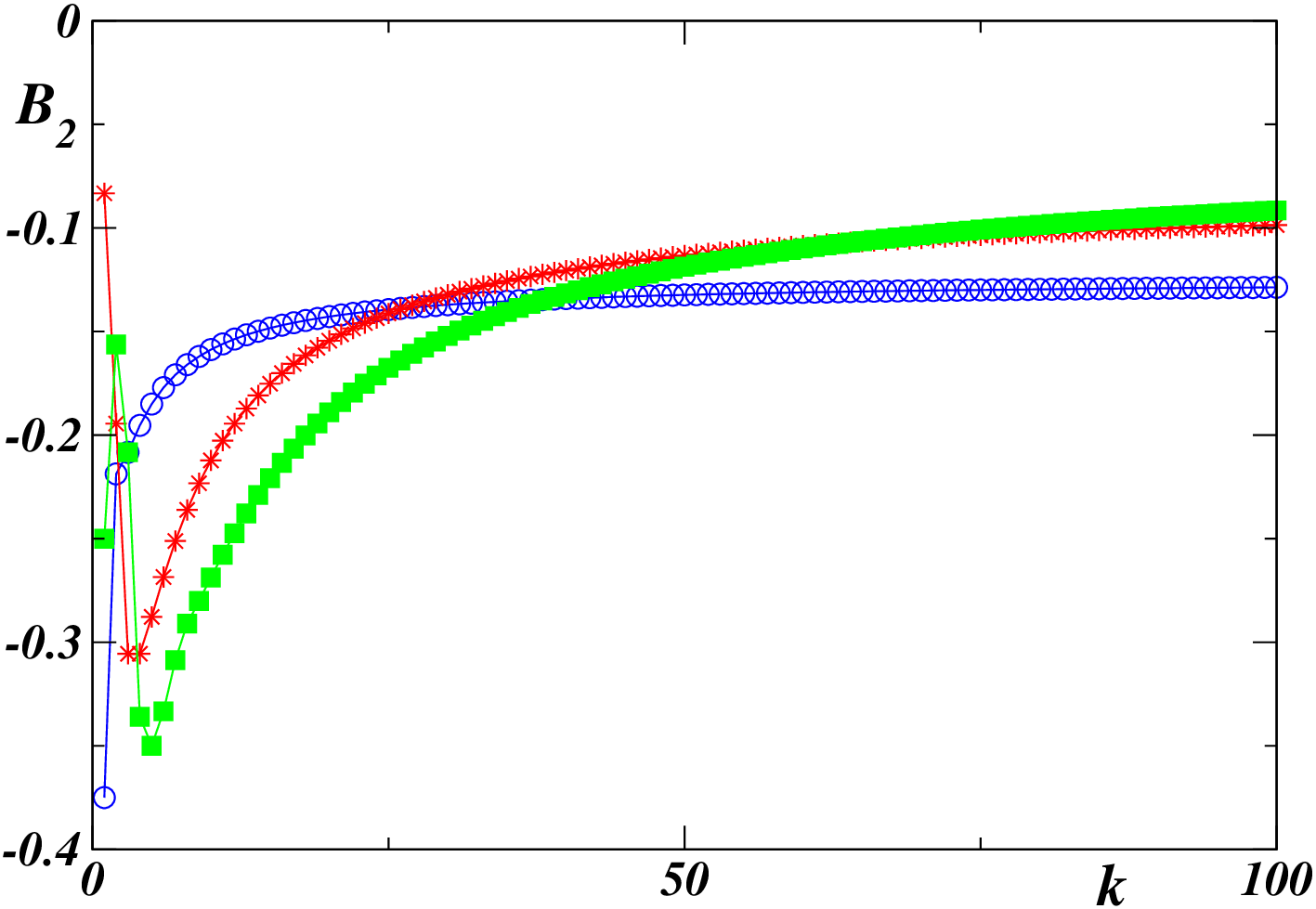}}}
%\vspace{0.5cm}
\caption{$B_2$ vs. $k=4\pi \kappa$ for an hard-core NACS gas with $l=1/2$ (blue open circles), $l=1$ (red stars), 
$l=3/2$ (green squares). In this and the following 
figures $B_2$ is in units of $\lambda_T^2$ (furthermore the line 
is just a guide for the eye, since $k$ assumes only integer 
values). The values of $B_2$ in the limit 
$\kappa\rightarrow \infty$ are given by 
$-\frac{1}{8},-\frac{1}{12},-\frac{1}{16}$ 
and are correctly reproduced.}
\label{fig3}
\end{figure}

\begin{figure}[t]
\centerline{\scalebox{0.40}{\includegraphics{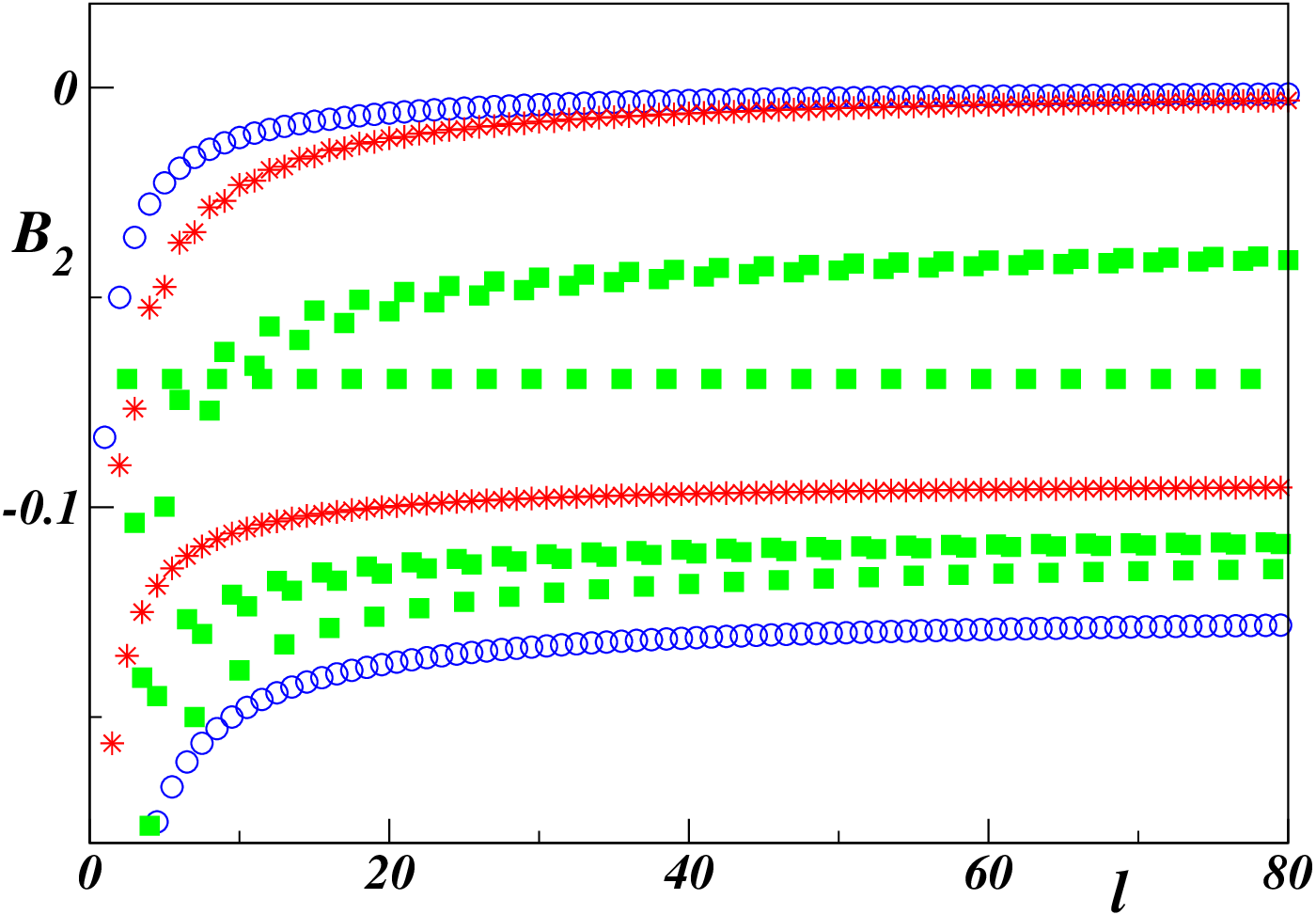}}}
%\vspace{0.5cm}
\caption{$B_2$ vs. $l$ for an hard-core NACS gas with 
$k=1$ (blue open circles), $k=2$ (red stars), $k=3$ (green squares). 
$l$ varies over all the integer and the half-integer numbers: 
in the upper (lower) part of the figure the plotted values 
of $B_2$ correspond to integer (half-integer) values of $l$.}
\label{fig4}
\end{figure}

Eq. (\ref{coeffviriale}) is the main result of this Subsection: 
the dependence of $B_2$ on $k$ for some fixed $l$'s, and vice versa 
on $l$ for some fixed $k$'s is represented in Figs.\ref{fig3}-\ref{fig4}: 
one sees from Fig.\ref{fig3} a non-monotonic behavior of $B_2$ as a function 
of $k$. In Fig.\ref{fig4} one can see that the values of $B_2$ vs. $l$ 
form two different groups, depending on integer and half-integer values 
of $l$.

As a first check of Eq. (\ref{coeffviriale}), we observe that 
the value of $B_2^{(n.i.)}(l,T)$ of the free case (corresponding 
to the limit $1/4 \pi \kappa \rightarrow 0$) is correctly 
reproduced: indeed, for a given $l$ one has 
in this limit $\omega_j\rightarrow 0^{\pm}$, 
$\gamma_j \rightarrow \left\{\begin{array}{ll} 0^{+}\\2^{-}\end{array}\right.$, 
$\gamma_j^2- 2 \gamma_j\rightarrow 0^-$, $[(\gamma_j+1) \, mod \, 2-1 ]^2 
\rightarrow 0^{+}$, as it might. A more detailed discussion 
on the limit $1/4 \pi \kappa \rightarrow 0$ 
is presented in Appendix \ref{hagen_appendix}.
%By a careful inspection, it is possible to conclude that 
%Eq. (30) of \cite{Lee95} and Eq. (3) of \cite{Hagen96} 
%do not show on the contrary the correct behavior in the same limit. 
%, how shown through the plots in Fig.\ref{fig3} and Fig.\ref{fig4}. 

To further compare with available results, we observe 
that in \cite{Lee95} and \cite{Lee96} it was stated that the 
factors $(-1)^{2l}$ should not appear in the expression of the 
two-particle partition function, or in the expression of the second virial coefficient 
($l$ denoting the isospin quantum number of each particle). However, 
as pointed out in \cite{Hagen96} and as further motivated in the following, 
such factors are needed. To clarify this issue it is convenient to 
make reference to the properties of the Clebsch-Gordan coefficients. 
The two (spinless) particles in exam have both isospin $l$ and total 
isospin $j$: the Clebsch-Gordan coefficients express the change of basis, 
in the two-body isospin space, between the basis labeled by 
the individual magnetic isospin numbers $m_1$, $m_2$ and 
the basis labeled by the total and magnetic isospin $j,m_j$: 
$$\vert l l j m_j\rangle=\sum_{m_1=-l}^{l}\sum_{m_2=-l}^{l}\langle l m_1 l m_2\vert j m_j \rangle  \vert l m_1 l m_2 \rangle\,\,\,,$$ 
where the Clebsch-Gordan coefficients fulfill the symmetry property 
$$\langle l m_1 l m_2\vert j m_j \rangle=(-1)^{2l-j} \langle l m_2 l m_1\vert 
j m_j \rangle\,\,\,.$$ 
Notice that the real spin of the particles 
is not taken in account in this consideration. 
With respect to the exchange of all the quantum numbers, the isospin two-body 
wavefunction corresponding to a state of total isospin $j$ takes a factor 
$(-1)^{2l-j}=(-1)^{j+2l}$ (being $j$ an integer), 
so that the factor $\frac{1+(-1)^{j+2l}}{2}$ projects 
over the states for which the partition function can be evaluated 
in the bosonic basis, the factor $\frac{1-(-1)^{j+2l}}{2}$ projects over 
those for which the partition function can be evaluated in the fermionic 
basis, from which Eqs. (\ref{deltabidue}), (\ref{limitenoninteragente}) 
and (\ref{coeffviriale}) can be obtained.

Our result (\ref{coeffviriale}) differs also from the results presented 
in \cite{Lo93-2}, where a method of computation of the second virial 
coefficient for hard-core NACS gas based 
on the idea of averaging over all the isospin states 
is proposed. In particular, in \cite{Lo93-2} 
the special cases $l=1/2$, 
$l=1$, and the large-$\kappa$ limit for two particles belonging to 
a representation $l$ with $\lim_{l\rightarrow\infty}\frac{l^2}{4 \pi \kappa}=a<1$ 
were considered.   
In the last limit the sum over all the resulting total isospins  
$r\leq 2l$ is approximated by an integral. The results are given by 
Eqs.\,(35),(36) and (38) of \cite{Lo93-2}:
\begin{eqnarray}
&& B_2^{h.c.}\left(k,l=1/2,T\right) = \lambda_T^2 \left( -\frac{1}{4}+\frac{3}{4 k}-\frac{3}{8 k^2} \right)\,\,\,, \nonumber \\
&& B_2^{h.c.}\left(k,l=1,T\right) = 
\lambda_T^2 \left( -\frac{1}{4}+\frac{20}{9 k}-\frac{8}{3 k^2} \right) \quad \forall k>1\,\,\,, 
\label{Lo_Comp_1} \\  
&& 
\lim_{l\rightarrow\infty}B_2^{h.c.}\left(k,l,T\right)=\lambda_T^2 \left( -\frac{1}{4}+a-\frac{a^2}{3} \right)\,\,\,, \nonumber 
\end{eqnarray} 
while our corresponding results are
\begin{eqnarray}
&& B_2^{h.c.}\left(k,l=1/2,T\right) = \lambda_T^2 \left( -\frac{1}{8}+\frac{3}{8 k}-\frac{3}{8 k^2} \right)\quad \forall k\geq 2\,\,\,, \nonumber \\ 
&& B_2^{h.c.}\left(k,l=1,T\right) = \lambda_T^2 \left( -\frac{1}{12}+\frac{14}{9 k}-\frac{8}{3 k^2} \right) \quad \forall k\geq 4 \,\,\,, \label{Lo_Comp_2} \\
&& \lim_{l\rightarrow \infty}B_2^{h.c.}\left(k,l,T\right)=\lambda_T^2 \left( -\frac{a}{4}-\frac{2}{3}a^2 \right) \,\,\,.  \nonumber 
\end{eqnarray}
The limits $l \rightarrow \infty$ in (\ref{Lo_Comp_1}) and (\ref{Lo_Comp_2}) 
are taken together with 
$\lim_{l\rightarrow\infty} l^2 / k=a$, where $a$ is kept fixed and 
$a<1$ in (\ref{Lo_Comp_1}) and $a<1/2$ in (\ref{Lo_Comp_2}). 
The derivation from Eq. (\ref{coeffviriale}) of the three special cases above is reported in the Appendix \ref{appendix_lo}. 
We notice that the asymptotic value found for $B_2$ in the third case 
is expected to vanish for $a=0$, while on the contrary  
this does not occur for the results of \cite{Lo93-2}: indeed, 
$a=0$ corresponds to consider the limit 
$B_2^{(n.i.)}(l,T)\,=\, \frac{-1}{2l+1}\frac{\lambda^2_T}{4}$, 
which vanishes in the large $l$-limit. 

The difference between the results of Ref. \cite{Lo93-2} and ours 
stands in a different averaging: while in \cite{Lo93-2}  the virial 
coefficients 
are expressed as averages of the virial coefficients over the 
$(2l+1)^2$ two-body states of isospin, in our case we take into account 
the effect of the isospin symmetry factor $(-1)^{j+2l}$ 
characterizing the states of total isospin $j.$  

Notice that Eq. (\ref{coeffviriale}) for the second virial coefficient 
can be recovered using the approach presented in \cite{Hagen96}, as 
shown in Appendix \ref{hagen_appendix}. Indeed, one can find from 
Eqs. (\ref{legamepartizioneviriale})-(\ref{limitenoninteragente})-(\ref{coeffviriale})(see 
Appendix \ref{hagen_appendix} for details) that the following expression for 
$B_2^{h.c.}$ given in Eq. (2) of \cite{Hagen96} holds:
\begin{equation} B_2^{h.c.}(\kappa,l, T) = 
\frac{1}{(2l +1)^2} \sum^{2l}_{j=0} (2j+1) 
\left[\frac{1+(-1)^{j+2l}}{2} B^B_2(\omega_j,T) + 
\frac{1-(-1)^{j+2l}}{2}B^F_2(\omega_j,T)\right]\,\,\,, \label{Hagen}
\end{equation}
where $B^{B,F}_2(\omega,T)$ is given by \cite{Arovas85} 
\begin{equation} B^{B(F)}_2(\omega,T) = 
\frac{1}{4}\lambda^2_T \left\{ \begin{array}{ll} 
                 -1 + 4\delta - 2\delta^2, & \mbox{N even (odd)} \\ 
                 1 - 2\delta^2, & \mbox{N odd (even)} 
                 \end{array} 
                 \right. \label{famousresult}\end{equation} 
($\omega=N+\delta$ and $N$ an integer such that $0 \leq \delta < 1$). 
The computation 
presented in Appendix \ref{hagen_appendix} shows that Eq. (2) of \cite{Hagen96}
is a correct starting point to study $B_2^{h.c.}$: however, notice that 
Eq. (3) of \cite{Hagen96} should be replaced with 
Eq. (\ref{trecorretta}) given in 
Appendix \ref{hagen_appendix}.

We finally discuss in more detail the non-interacting limit 
$1 / 4\pi\kappa \rightarrow 0$ in order to clarify the meaning of Eq. (\ref{limitenoninteragente}). 
In the limit $k \rightarrow \infty$ the covariant derivatives in 
(\ref{hamiltonianadelmodello}) trivialize and 
the isospin becomes just a symmetry of the Hamiltonian, 
resulting in a pure (isospin) degeneration $g=2l+1$. 
Let $\Xi$ be the grand partition function, 
$\epsilon$ the generic single-particle energy level for an assigned spectral discretization, and using the upper/lower signs 
respectively for the $g$-degenerate bosonic/fermionic single-particle states: 
notice that any value of $g$ is allowed both in the bosonic and 
the fermionic case, since the statistics is not constrained by isospin. 
The expressions for the grand partition function, the 
pressure and the density are 
$$\Xi(z,A,T)=\prod_{\epsilon} (1 \mp z e^{-\beta \epsilon})^{\mp g}\,\,\,,$$ 
$$\frac{PA}{k_B T}=\ln \Xi(z,A,T)=\mp \sum_\epsilon g \ln(1 \mp z 
e^{-\beta \epsilon})$$ 
and 
$$ N\equiv z \frac{\partial }{\partial z} \ln \Xi(z,A,T)= g \sum_{\epsilon}\frac{1}{z^{-1} e^{\beta \epsilon}\mp 1}\,\,\,.$$  
It follows 
\begin{equation} \left\{ \begin{array}{l} 
\frac{(P/g) A}{k_B T}=\mp \sum_\epsilon \ln(1 \mp z e^{-\beta \epsilon})  \\ 
(\rho/g)= \sum_{\epsilon}\frac{1/A}{z^{-1} e^{\beta \epsilon}\mp 1} \end{array}\right. 
\Rightarrow \,\,\frac{P}{g}= k_B T \sum_{n=0}^{\infty} \left(\frac{\rho}{g} 
\right)^n B_n^o\,\,\,,\end{equation} where $B_n^o$ denotes the $n$-th virial coefficient for spinless boson(/spinless fermion) 
without either isospin degeneration. Hence, denoting by $B_n$ 
the $n$-th virial coefficient in presence of isospin freedom one has 
$$P=k_B T \sum_{n=0}^{\infty} \left(\rho \lambda_T^2\right)^n \frac{1}{g^{n-1}} B_n^0 =k_B T \sum_{n=0}^{\infty} \rho^n B_n $$ 
and therefore
$$
B_n= \frac{1}{g^{n-1}} B_n^0\,\,\,:$$  
therefore $B_2= \frac{1}{2l+1} B_2^0$, that is exactly what is written in Eq. 
(\ref{limitenoninteragente}). The result (\ref{limitenoninteragente}) can be also understood by observing 
that all the virial coefficients for a system of NACS defined over a representation of isospin $l$ tend, 
in the non-interacting limit $\kappa \rightarrow \infty$, to $(-1)^{2l}$ times those of an ideal gas of identical 
$l$-spin ordinary quantum particles. In particular, the second virial coefficient of an ideal system of quantum $s$-spin particles is indeed:  
\begin{equation} 
B_2(s,T)  = +\frac{\lambda^2_T}{4}  \frac{(-1)^{2s+1}}{2s+1}\,\,\,,
\label{gasidealiconspin}
\end{equation}
in agreement with (\ref{limitenoninteragente}). The issue becomes much more complex for flux-carrying particles (finite $\kappa$) having a non-zero spin, 
as discussed in \cite{Hagen90,Blum90,Horvathy05}: however, for the true ideal spinor case $\alpha=0$ and $s=1/2$ it is $B_2(s=1/2, T)=\frac{1}{8}\lambda_T^2$ \cite{Blum90}, 
again in agreement with (\ref{gasidealiconspin}) and the related (\ref{limitenoninteragente}).

We conclude this Section by observing that a semiclassical computation of the second virial coefficient for a system of hard-core NACS particles 
reproduces Eq. (\ref{coeffviriale}): we remind that for an Abelian hard-core gas the semiclassical approximation \cite{Bhaduri91,Khare05} 
yields the exact quantum result of \cite{Arovas85} for $B_2$. 
By extending such a computation 
to the hard-core NACS gas we find exactly Eq. (\ref{coeffviriale}) (details are not reported here). We mention that 
in literature it has been conjectured that the semiclassical approximation could give the exact expressions for all the virial coefficients in presence of hard-core boundary 
conditions \cite{Khare05-2}: the rationale for this conjecture is that for hard-core boundary conditions there are no other length scale besides $\lambda_T$. 
Therefore, having established the extension to the non-Abelian hard-core case of the semiclassical computation of $B_2$, 
one could in the future obtain information 
about higher virial coefficients for the hard-core NACS gas. 
However, we alert the reader that the presence of other relevant length scales (other than $\lambda_T$) in general prevent the semiclassical approximation 
from being exact: an explicit example is given in \cite{Khare05-2}. We conclude that for the soft-core NACS (that we are going to treat in 
the next Section) the semiclassical approximation is not expected 
to give the correct results.

\subsection{General Soft-Core Case}\label{generalsoftcorecase} 
If one removes the hard-core boundary condition for the 
relative $(2l+1)^2$-component two-anyonic wavefunction and 
fixes an arbitrary external potential as a spectral regularizator, 
then the spectrum of each projected Hamiltonian operator $H_j'$ 
can be represented as the union of the spectra of $(2j+1)$ scalar 
Schr\"odinger operators, one for each $j_z$-component, 
endowed with its respective hard-core parameter 
$\varepsilon_{j,j_z}$ (as shown in Appendix \ref{spectrum_sc}). 
As discussed in Section \ref{nonabeliananyons}, 
one then ends up with a set of $(2l+1)^2$ (in principle independent) 
parameters $\varepsilon_{j,j_z}$, which are needed to fix the boundary behavior. 
They can be organized in a $(2l+1)\times (2l +1)$ matrix: 
\begin{equation} 
\left( 
\begin{array}{lllll} 
\varepsilon_{0,0} & \varepsilon_{1,1} & \varepsilon_{2,2} & \cdots & \varepsilon_{2l+1,2l+1}\\ 
\varepsilon_{1,-1} & \varepsilon_{1,0} & \varepsilon_{2,1} & \cdots & \varepsilon_{2l+1,2l}\\ 
\varepsilon_{2,-2} & \varepsilon_{2,-1} & \varepsilon_{2,0} & \cdots & \cdots \\ 
\cdots & \cdots & \cdots & \cdots & \cdots \\ 
\varepsilon_{2l+1,-2l-1} & \varepsilon_{2l+1,-2l} & \cdots & \cdots & \varepsilon_{2l+1,0} \\ 
\end{array} 
\right)\,\,\,. 
\label{matrix}
\end{equation} 
Proceeding as in the previous Subsection \ref{hardcorecase}, one has then
for the general soft-core NACS gas the following expression for the 
second virial coefficient:
\begin{equation} 
B_2^{s.c.}\left( \kappa,l,T \right)=
\frac{1}{(2l+1)^2}\sum_{j=0}^{2l}\sum_{j_z=-j}^{j}\left[\frac{1+(-1)^{j+2l}}{2} B^B_2(\omega_j,T,\varepsilon_{j,j_z}) + 
\frac{1-(-1)^{j+2l}}{2}B^F_2(\omega_j,T,\varepsilon_{j,j_z})\right]\,\,\,,
\label{generalformulasoftcorecase} \end{equation} 
where $B^B_2(\omega_j,T,\varepsilon_{j,j_z})$ is the soft-core expression entering Eq. (\ref{explicitintegralform}):
\begin{equation} 
B^B_2(\omega_j,T,\varepsilon_{j,j_z})=B_2^{h.c.}(\delta_j,T)- 2 \lambda_T^2 
\left\{ 
e^{\varepsilon_{j,j_z}} \theta(-\sigma)  
+\frac{\delta_j\sigma}{\pi} \left(\sin{\pi \delta_j} \right)
\int_0^\infty \frac{dt e^{-\varepsilon_{j,j_z}\, t} t^{\vert\delta_j\vert-1}}{1+2\sigma(\cos{\pi\delta_j})\;t^{|\delta_j|}+t^{2|\delta_j|}} \right\},
\end{equation}
with $\delta_j\equiv (\omega_j+1) \, mod \, 2-1$, and 
$B^F_2(\omega_j,T,\varepsilon_{j,j_z})$ is the previous expression 
evaluated for $\omega_j \rightarrow \omega_j+1$:   
\begin{equation} 
B^F_2(\omega_j,T,\varepsilon_{j,j_z})=B_2^{h.c.}(\Gamma_j,T)- 2 \lambda_T^2 
\left\{ 
e^{\varepsilon_{j,j_z}} \theta(-\sigma)  
+\frac{\Gamma_j \,\sigma}{\pi} \left(\sin{\pi \Gamma_j} \right)
\int_0^\infty \frac{dt e^{-\varepsilon_{j,j_z}\, t} t^{|\Gamma_j|-1}}{1+2\sigma(\cos{\pi\Gamma_j})\;t^{|\Gamma_j|}+t^{2|\Gamma_j|}} \right\},
\end{equation}
with $\Gamma_j\equiv \omega_j \, mod \, 2\,-1$. 
Eq. (\ref{generalformulasoftcorecase}) is the desired result 
for a NACS ideal gas with general soft-core boundary conditions.

\begin{figure}[t]
\centerline{\scalebox{0.40}{\includegraphics{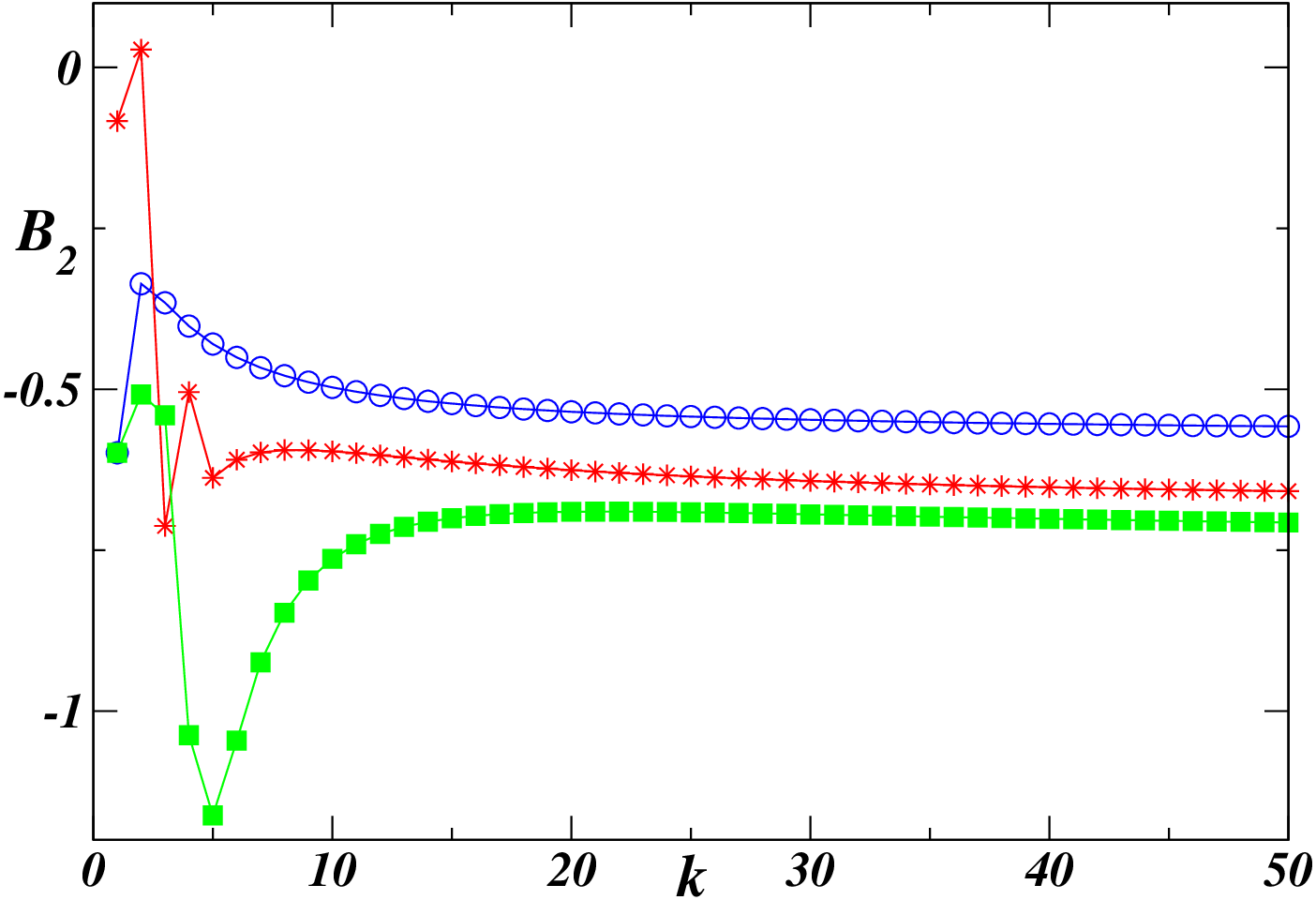}}}
%\vspace{0.5cm}
\caption{$B_2^{s.c.}(k,l,T)$ vs. $k$ for a soft-core NACS gas with $l=1/2$ 
(blue open circles), $l=1$ (red stars) and $l=3/2$ (green squares): 
in all $\varepsilon=0.1$ ($k$ varies over the positive integers).}
\label{fig5}
\end{figure}

\begin{figure}[t]
\centerline{\scalebox{0.40}{\includegraphics{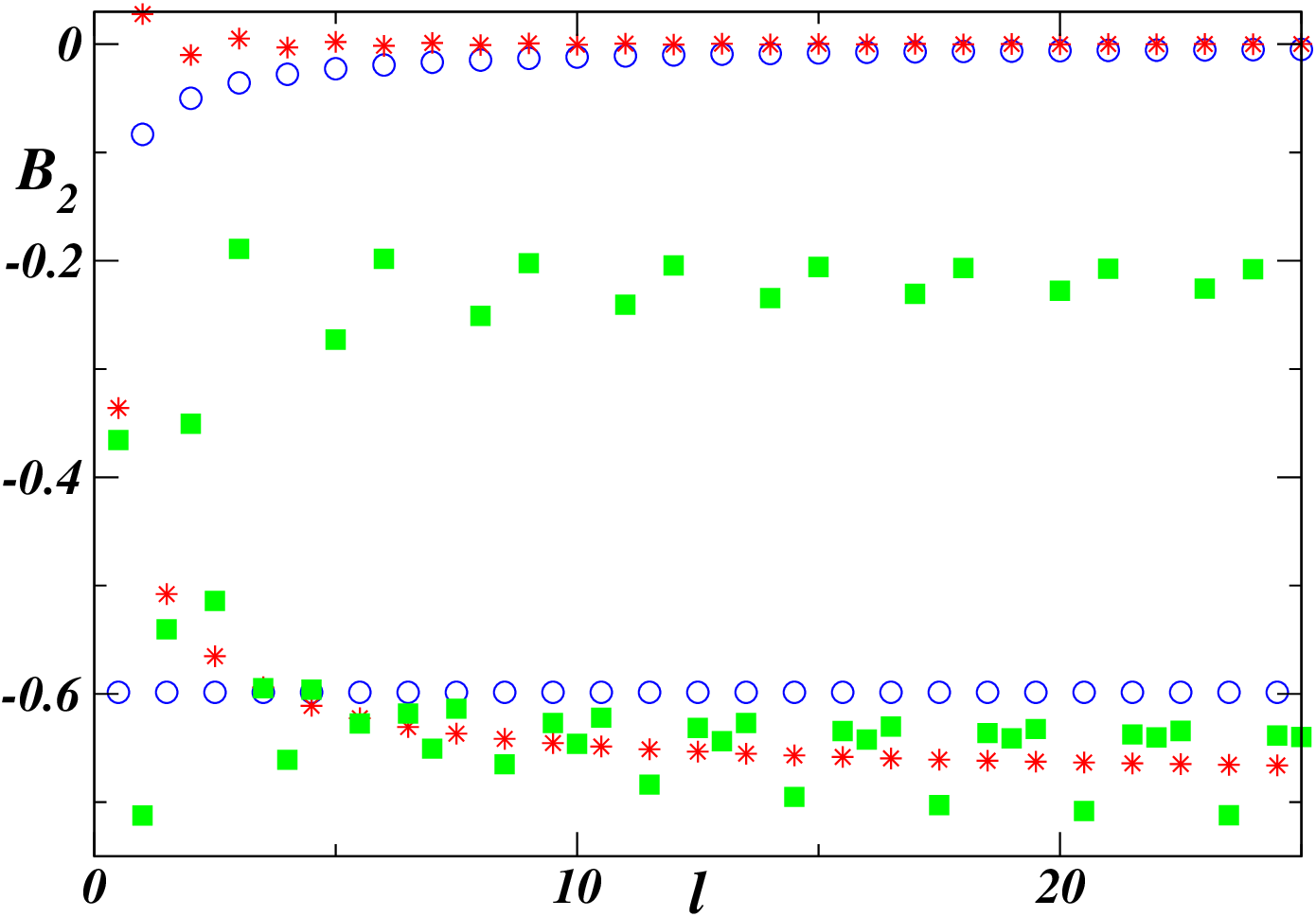}}}
%\vspace{0.5cm}
\caption{$B_2^{s.c.}(k,l,T)$ vs. $l$ for $k=1$ (blue open circles), 
$k=2$ (red stars) and $k=3$ (green squares), with $\varepsilon=0.1$ 
($l$ varies over the integer and the half-integer numbers).}
\label{fig6}
\end{figure}

\begin{figure}[t]
\centerline{\scalebox{0.40}{\includegraphics{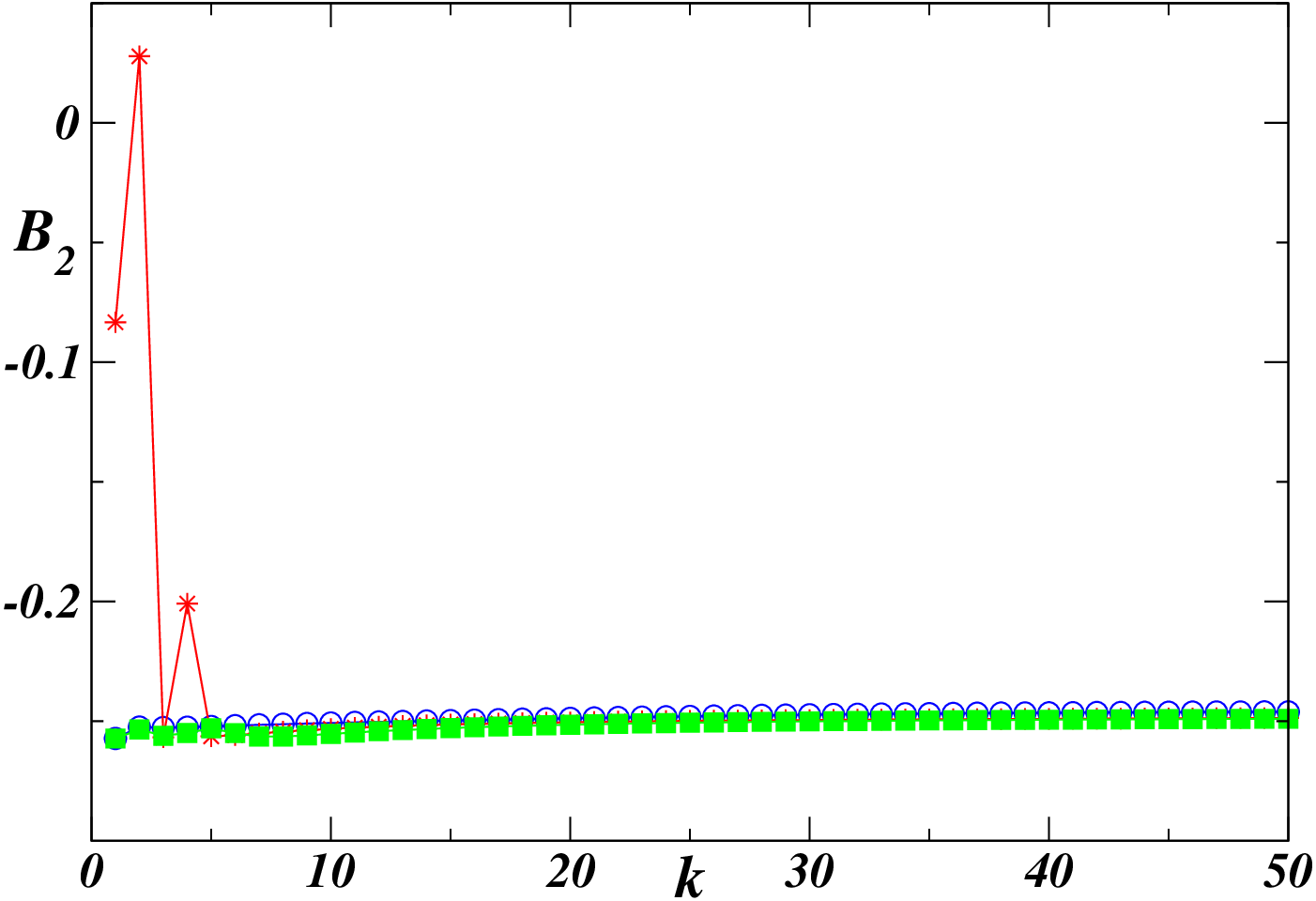}}}
%\vspace{0.5cm}
\caption{$B_2^{s.c.}(k,l,T)$ vs. $k$ for $l=1/2$ 
(blue open circles), $l=1$ (red stars) and $l=3/2$ (green squares), with $\varepsilon=1.4$.}
\label{fig7}
\end{figure}

\begin{figure}[t]
%\vspace{0.7cm}
\centerline{\scalebox{0.40}{\includegraphics{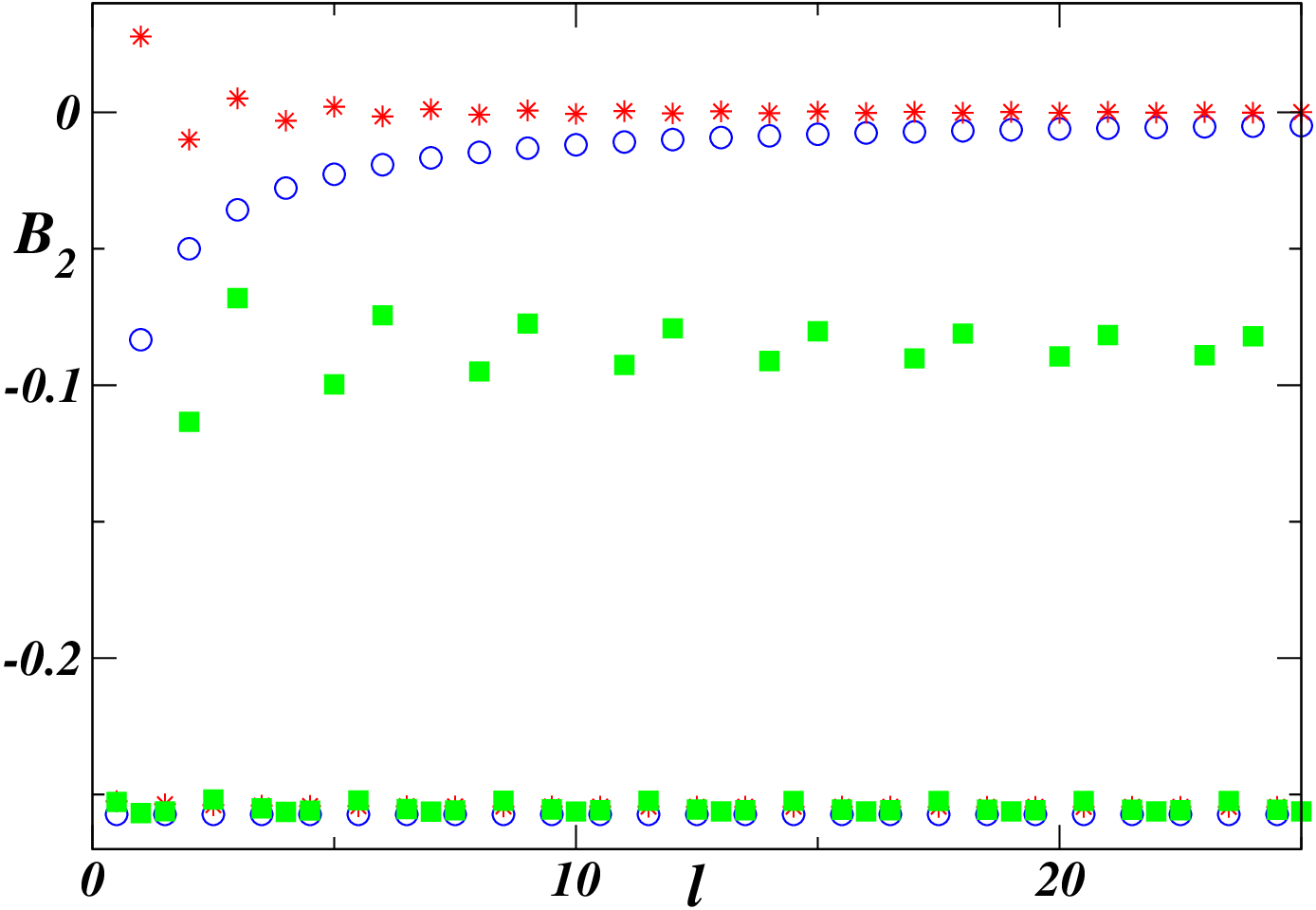}}}
%\vspace{0.5cm}
\caption{$B_2^{s.c.}(k,l,T)$ vs. $l$ for $k=1$ (blue open circles), 
$k=2$ (red stars) and $k=3$ (green squares), with $\varepsilon=1.4$.}
\label{fig8}
\end{figure}

\begin{figure}[t]
\centerline{\scalebox{0.40}{\includegraphics{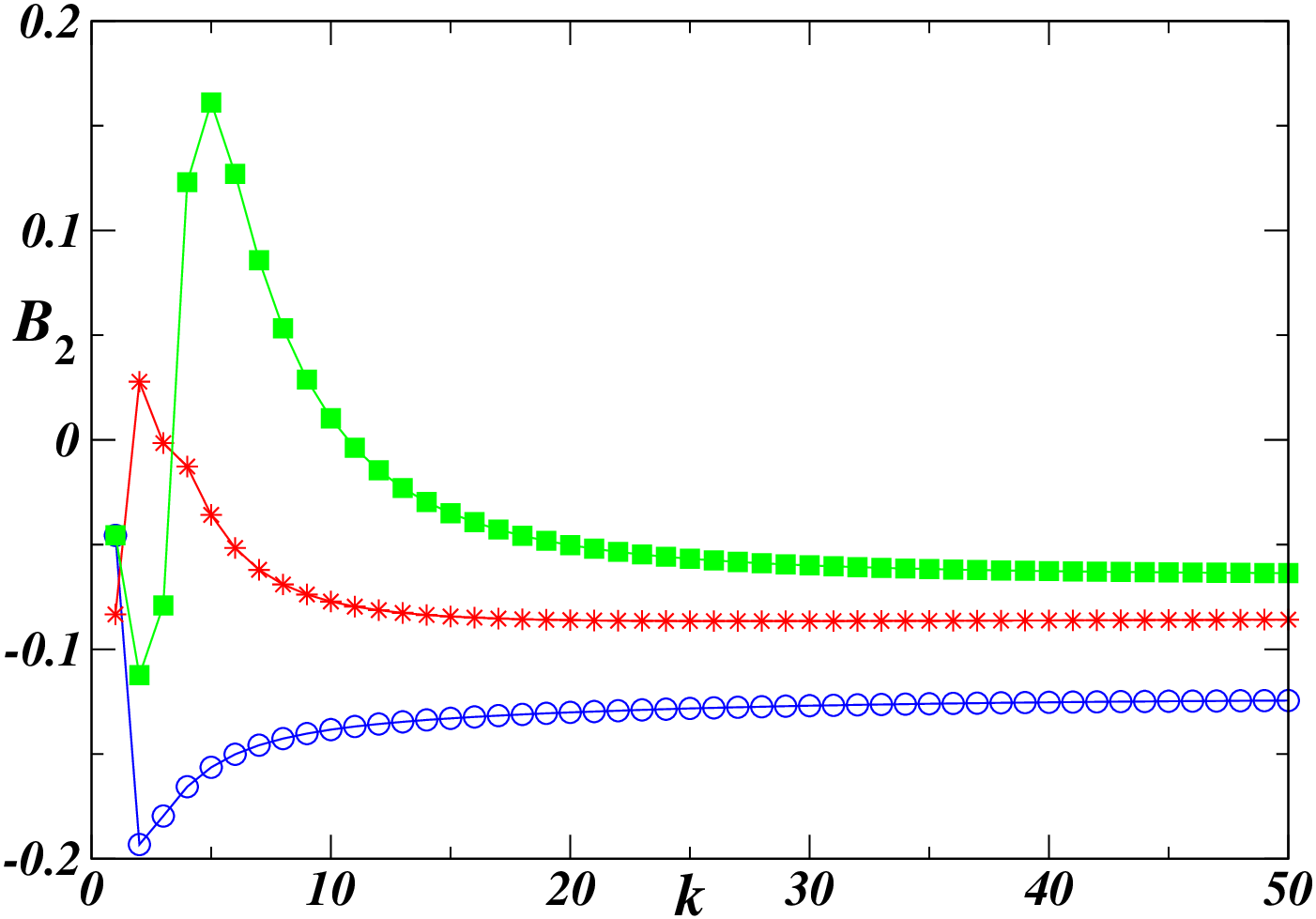}}}
%\vspace{0.5cm}
\caption{$B_2^{s.c.}(k,l,T)$ vs. $k$ for $l=1/2$ 
(blue open circles), $l=1$ (red stars) and $l=3/2$ (green squares), with $\varepsilon=10$.}
\label{fig9}
\end{figure}

\begin{figure}[t]
%\vspace{0.7cm}
\centerline{\scalebox{0.40}{\includegraphics{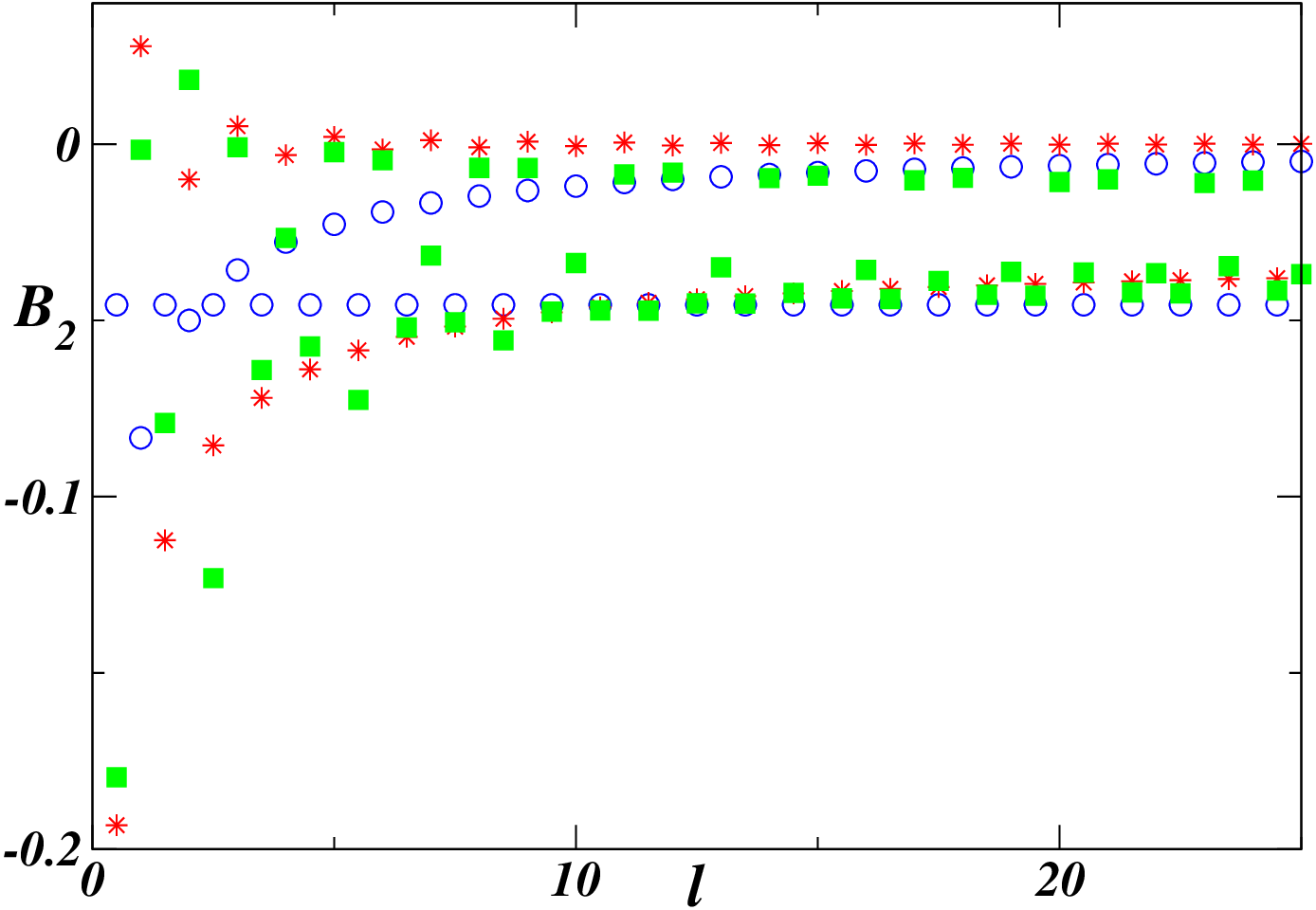}}}
%\vspace{0.5cm}
\caption{$B_2^{s.c.}(k,l,T)$ vs. $l$ for $k=1$ (blue open circles), 
$k=2$ (red stars) and $k=3$ (green squares), with $\varepsilon=10$.}
\label{fig10}
\end{figure}

To perform explicit computations, we consider in the following 
the simple case in which the isotropy of the hard-core parameter 
is assumed within each shell with assigned isospin quantum number $l$.  
In other words, $\varepsilon_{j,j_z}\equiv\varepsilon_j$ and the matrix 
(\ref{matrix}) then reads
\begin{equation} 
\varepsilon_{j,j_z}\equiv\left( 
\begin{array}{llll} 
\varepsilon_{0} & \varepsilon_{1} & \cdots & \varepsilon_{2l+1}\\ 
\varepsilon_{1} & \varepsilon_{1} & \cdots & \varepsilon_{2l+1}\\ 
\cdots & \cdots & \cdots & \cdots \\ 
\varepsilon_{2l+1} & \varepsilon_{2l+1} & \cdots  & \varepsilon_{2l+1} \\ 
\end{array} 
\right)\,\,\,. 
\label{matrix_is}
\end{equation} 
When all the element of the matrix (\ref{matrix_is}) are equal, we will 
use the notation $\varepsilon_{j,j_z} \equiv \varepsilon$. 
In such a completely isotropic case, 
Eq. (\ref{generalformulasoftcorecase}) takes the 
simplified form
\begin{equation} 
\label{simplifiedsoftcore}
B_2^{s.c.}\left(\kappa,l,T\right)=
\frac{1}{(2l+1)^2}\sum_{j=0}^{2l}(2j+1)\,B^B_2(\nu_j,T,\varepsilon)\,\,\,,
\end{equation}
where 
\begin{equation}
\label{modifiedparameter}
\nu_j\equiv\left(\omega_j-\frac{1+(-1)^{j+2l}}{2}\right) \, mod \, 2-1\,\,\,.\\
\end{equation}
In Figs.\,\ref{fig5}-\ref{fig10} we show, for three 
values of the isotropic hard-core parameter $\varepsilon$, the dependence of $B_2^{s.c.}$ 
on $k$ for some fixed $l$'s, and vice versa on $l$ for some fixed $k$'s. 
Fig.\ref{fig7} evidences that for suitable values of $\varepsilon$ 
the second virial coefficient may change sign and have strong variations. 
From Eq. (\ref{simplifiedsoftcore}) it is possible to see that 
the values of $B_2^{s.c.}\left(\kappa,l,T\right)$ 
corresponding to semi-integer $l$ and $k=1$ are independent of $l$, 
depending only on $\varepsilon$ and $T$ 
(see Figs.\ref{fig6},\ref{fig8},\ref{fig10}). 
In fact Eq. (\ref{modifiedparameter}) yields that for 
$l$ semi-integer ($l=1/2,3/2,\cdots$) and $k=1$ 
one has $\nu_j=\pm \frac{1}{2}$ and therefore 
\begin{equation} 
\label{simplifiedsoftcore_l}
B_2^{s.c.}\left(k=1,l=n+\frac{1}{2},T\right)=
B^B_2\left(\frac{1}{2},T,\varepsilon\right)\,\,\,
\end{equation}
(with $n=0,1,2,\cdots$).

For $l=1/2$, i.e. the lowest possible value of $l$ for non-Abelian anyons, 
the assumption of isotropy ($\varepsilon_{0,0}=\varepsilon_0$ 
and $\varepsilon_{1,m}=\varepsilon_1$ with $m=1,0,-1$) yields: 
\begin{equation} 
B_2^{s.c.}\left(\kappa,l=\frac{1}{2},T\right)=\,\frac{3}{4}B^B_2(\omega_1,T,\varepsilon_{1}) +\frac{1}{4}B^F_2(\omega_0,T,\varepsilon_{0})\,\,\,. 
\label{isotropictwodimensional} 
\end{equation} 
As example, let us consider the case $l=1/2, 4 \pi \kappa=3$:    
\begin{equation} 
B_2^{s.c.}\left(k=3,l=\frac{1}{2},T\right)=\,\frac{3}{4}B^B_2
\left(\alpha=\frac{1}{6},T,\varepsilon_{1}\right) + 
\frac{1}{4}B^F_2\left(\alpha=-\frac{1}{2},T,\varepsilon_{0}\right) 
\label{esempio} 
\end{equation} 
(similar results can be found for other values of $k$). 

\begin{figure}[t]
\vspace{0.4cm}
\centerline{\scalebox{0.35}{\includegraphics{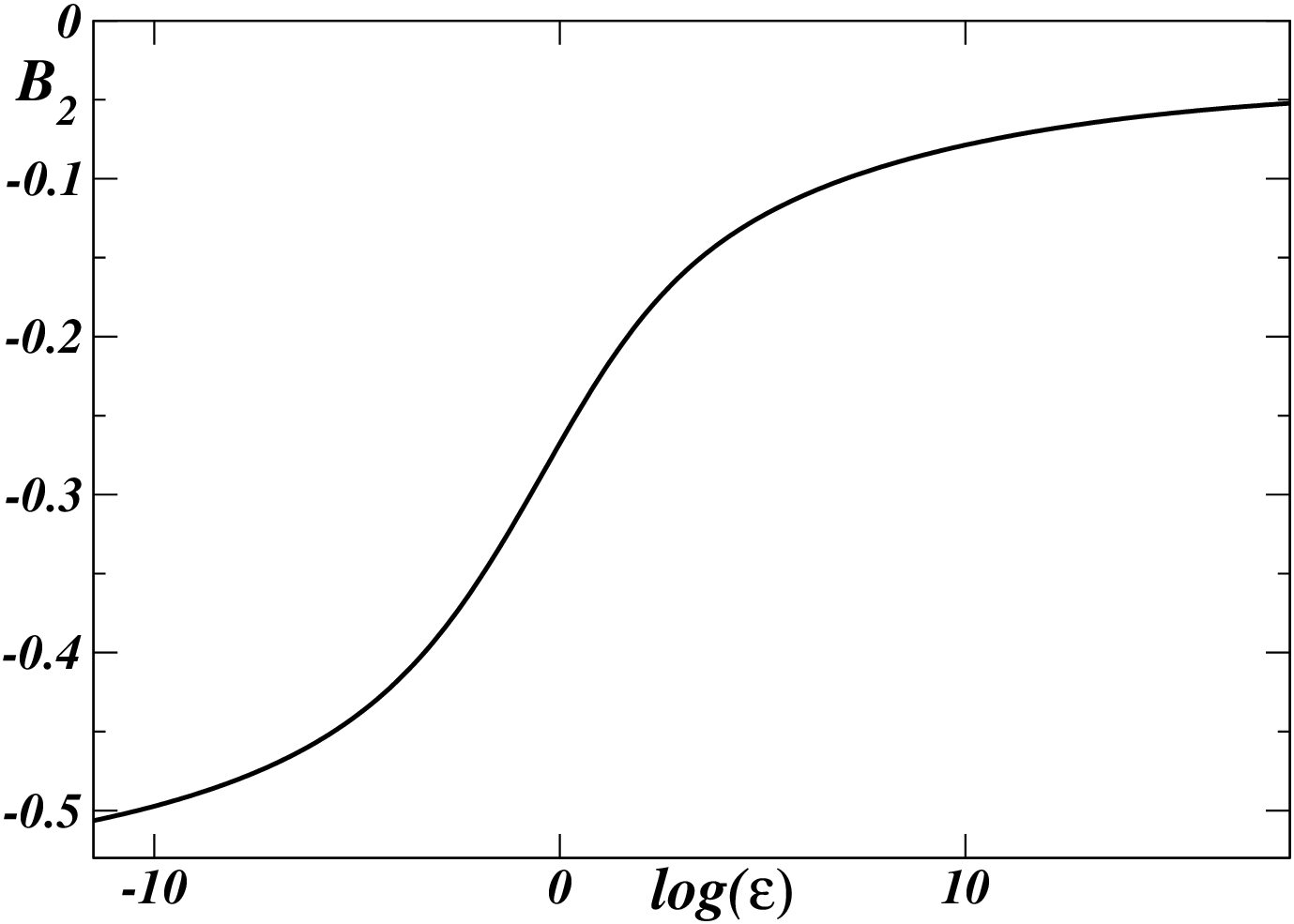}}}
%\vspace{0.5cm}
\caption{$B_2^{s.c.}$ as a function of the hard-core parameter $\varepsilon$ in logarithmic scale, for $k=3$ and $l=1/2$, in the completely isotropic case 
($\varepsilon_{0,0}=\varepsilon_{1,m}=\varepsilon$, for $m=1,0,-1$).}
\label{fig11}
\end{figure}

For the isotropic soft-core system, special boundary conditions are those limited to the $s$-channel (for which the $p$-wave is assumed to be hard-core, $\varepsilon_1=\infty$) and to the $p$-channel (for which, vice versa, the $s$-wave is assumed to be hard-core, $\varepsilon_0=\infty$). In order to assure the physical soundness of the virial expansion, $k_B\, T$ has to be much higher \cite{Giacconi96} than the energy of the eventual bound state $E_B$ associated to the wavefunction (\ref{boundstate}). Hence, for both these channels the virial expansion is meaningful provided that we take $\sigma=+1$ 
in Eq. (\ref{explicitintegralform}), 
and the virial coefficients for these two channels are   
\begin{equation} 
B_2^{s.c.}\left(k=3,l=\frac{1}{2}\right)_{s-\text{channel}}=
-\frac{\lambda_T^2}{24}\left\{1+\frac{24}{\pi}\int_0^\infty dt \frac{ e^{-\varepsilon_0 t} t^{-1/2}}{1+t}\right\} 
\end{equation} 
and 
\begin{equation} 
B_2^{s.c.}\left(k=3,l=\frac{1}{2}\right)_{p-\text{channel}}=-\frac{\lambda_T^2}{24}\left\{1+\frac{4}{\pi}\int_0^\infty dt \frac{ e^{-\varepsilon_1 t} t^{-5/6}}{1+\sqrt{3}\,t^{1/6}+t^{1/3}}\right\}\,\,\,. 
\end{equation} 
The previous equation clearly shows that the depletion of $B_2$ with 
respect to the hard-core value $-\frac{1}{24}\lambda_T^2$ 
is the result of the anyonic collisions allowed by the soft-core conditions.  
If the four parameters of the whole matrix are taken to be identical 
$\varepsilon_0 =\varepsilon_1\equiv\varepsilon$ 
("complete isotropy" of the hard-core parameters matrix), 
the expression for the virial coefficient reduces to  
\begin{equation} 
B_2^{(s.c.)}\left(k=3,l=\frac{1}{2},T\right)=-\frac{\lambda_T^2}{24}\left\{1+\frac{4}{\pi}\int_0^\infty dt\; e^{-\varepsilon t} \left(\frac{6 \; t^{-1/2}}{1+t}+\frac{t^{-5/6}}{1+\sqrt{3}\,t^{1/6}+t^{1/3}}\right) \right\}\,\,\,. 
\label{fiboisotro} 
\end{equation} 
The dependence of this quantity on $\varepsilon$ becomes more evident by representing the $\varepsilon$ variable in logarithmic scale, as shown in 
Fig.\ref{fig11}. The hard-core limit value  $B_2^{h.c.}/\lambda_T^2=-1/24$ predicted by (\ref{coeffviriale}) is asymptotically approached, although 
for extremely high $\varepsilon$: e.g. for $\varepsilon=10^{17}$ it 
is $B_2^{h.c.}/\lambda_T^2 \approx -0.05$, which deviates 
from the asymptotic value by a $\approx 20\%$. We conclude that even an extremely small deviation from the hard-core conditions may have a significant impact 
on $B_2$ and therefore on the thermodynamical properties. 
At variance, for small values of $\varepsilon$, the extension of the 
analysis presented in \cite{Kim98} for soft-core Abelian anyons 
allows to compute the value of $B_2$ 
for the limit case $\varepsilon=0$: for $\varepsilon=0$ one gets $B_2^{s.c.}\left(k=3,l=\frac{1}{2},T\right)/\lambda_T^2=-13/24$. 
The monotonically increasing behaviour of $B_2$ in $\varepsilon$ is evident from (\ref{fiboisotro}), 
and consistent with an approach towards an hard-core (hence more repulsive) condition.  

\section{Other Thermodynamical Properties}
\label{otherthermod} 
In this Section we remind the virial expansions for the main thermodynamical quantities of the system studied in this paper, 
namely the gas of NACS particles endowed with general boundary conditions for the wavefunctions and we discuss how these quantities 
read at the order $\rho \lambda_T^2$ of the virial coefficient, in order to highlight the role played by the second virial coefficient 
$B_2$ computed in the previous Section.

The thermodynamical quantities are associated to the virial coefficients $\{ B_n(T) \}$ of the equation of state 
(featuring the expansion of the pressure in powers of the number 
density $\rho$). 
As discussed in statistical mechanics textbooks \cite{Mayer77,Huang87,McQuarrie00} one has the following virial expansions 
for the pressure $P$, the Helmholtz free energy $A_H$, the Gibbs free energy $G$, the entropy $S$, the internal energy $E$ 
and the enthalpy $H$ ($A$ being the area) :
$$Pressure: \,\,\,\,\,\,\frac{PA}{Nk_BT}=1+\sum_{k \geq 1} B_{k+1}\, \rho^k\,\,\,;$$ 
$$Helmholtz \,\, free \,\, energy:\,\,\,\,\,\, \frac{A_H}{Nk_BT}=\log(\rho\lambda_T^2)-1+\sum_{k \geq 1} \frac{1}{k}\, B_{k+1}\, \rho^k\,\,\,;$$ 
$$Gibbs \,\, free \,\,energy:\,\,\,\,\,\, \frac{G}{Nk_BT}=\log(\rho\lambda_T^2)+\sum_{k \geq 1} \frac{k+1}{k}\, B_{k+1}\, \rho^k\,\,\,;$$ 
$$Entropy: \,\,\,\,\,\, \frac{S}{Nk_B}=2-\log(\rho\lambda_T^2)-\sum_{k \geq 1} \frac{1}{k}\, \frac{\partial}{\partial T}\left(T\, B_{k+1}\right)\, \rho^k\,\,\,;$$ 
$$Internal \,\, energy: \,\,\,\,\,\, \frac{E}{Nk_BT}=1-T\,\sum_{k \geq 1} \frac{1}{k}\, \frac{\partial B_{k+1}}{\partial T}\, \rho^k\,\,\,;$$ 
$$Enthalpy: \,\,\,\,\,\, \frac{H}{Nk_BT}=2+\sum_{k \geq 1} \left( B_{k+1}-\frac{1}{k}\,T\,\frac{\partial B_{k+1}}{\partial T}\right)\, \rho^k\,\,\,.$$

Using the previous expressions, 
stopping at the lowest order of the virial coefficient 
$\rho$ (i.e. $\rho \lambda_T^2$) and 
using the fact that for a general NACS ideal gas 
one has $B_{2}(T)\propto T^{-1}$, one can obtain the thermodynamical quantities at the lowest order of the virial expansion: in particular we find 
$$\frac{PA}{Nk_BT}=1+ B_{2} \rho\,\,\,;$$
$$\frac{A_H}{Nk_BT}=\log(\rho\lambda_T^2)-1+ B_{2}\, \rho\,\,\,;$$ 
$$\frac{G}{Nk_BT}=\log(\rho\lambda_T^2)+2 B_{2}\, \rho\,\,\,;$$ 
$$\frac{E}{Nk_BT}=1+ B_2\, \rho\,\,\,;$$ 
$$\frac{H}{Nk_BT}=2+ 2B_2 \,\rho$$
(at the lowest order of virial expansion, the entropy and the heat capacity at constant volume do not depend on $B_2$). Using the expression 
of $B_2$ given by Eq. (\ref{generalformulasoftcorecase}) one can obtain the deviation of the various thermodynamical quantities from their ideal gas value.

An important consequence of the previous results is that at the order $\rho \lambda_T^2$ one finds
\begin{equation}
E = PA:
\label{EOS}
\end{equation}  
Eq. (\ref{EOS}) is an exact identity for 2D Bose and Fermi ideal gases, valid at all the orders of the virial expansion \cite{Khare05}. Similarly, for the soft-core NACS ideal gas, at the order $\rho \lambda_T^2$, one has $H = 2E$, which is also exact at all orders for 
2D Bose and Fermi ideal gases \cite{Khare05}. Further investigations on the higher virial coefficients are needed to ascertain 
if the equation of state (\ref{EOS}) is exact (at all orders) for a general soft-core NACS ideal gas.

\section{Conclusions}
\label{conclusions}
In this paper we studied the thermodynamical properties of an ideal gas of 
non-Abelian Chern-Simons particles considering the effect of general 
soft-core boundary conditions for the two-body  
wavefunction at zero distance: we determined and studied the second virial 
coefficient as a function of the coupling $\kappa$ and the (iso)spin 
$l$ for generic hard-core parameters. This class of anyons is an ideal benchmark 
for checking our results against known results and for exploring the consequences of 
soft-core conditions. In this respect,  the present study may be regarded as a first step in 
obtaining results applicable to more general classes of anyons. A discussion of the comparison 
of obtained findings with available results 
in literature for systems of non-Abelian hard-core Chern-Simons 
particles has been also supplied. We also found 
that a semiclassical computation 
of the second virial coefficient for hard-core non-Abelian Chern-Simons particles gives the correct result, extending 
in this way the corresponding result for Abelian hard-core anyons. We have also wrote down the expressions for the thermodynamical quantities 
at the lower order of the virial expansion, finding that at this order the relation between the internal energy 
and the pressure is the same found (exactly) for 2D Bose and Fermi ideal gases. Further studies on the higher virial coefficients are needed to establish 
the eventual validity of the obtained relation between the pressure and the internal energy for a general soft-core NACS ideal gas.

\vspace{3mm}
{\it Acknowledgements:} We would like to thank 
Enore Guadagnini, Michele Burrello, Andrea De Luca and Andrea Cappelli 
for inspiring and very valuable discussions. We also thank Carl R. Hagen for 
useful discussions. 
%This work is supported by ... . 

\appendix

\section{Computation of virial coefficients in special cases}
\label{appendix_lo} 
To perform the comparison with Ref.\cite{Hagen96}, we compute in the following 
the virial coefficients for a NACS gas in the hard-core limit in 
the special cases considered in \cite{Lo93-2}:\\\\ 
$\star$ Case $l=1/2$ \ \ \ (with $k \geq 2$):  
\begin{equation} 
B_2^{h.c.}\left(k,l=\frac{1}{2},T\right)=
-\frac{\lambda_T^2}{8}-\frac{\lambda_T^2}{2(2l+1)^2} \Biggl[   [(\gamma_0+1)\mod\;  2-1]^2 + 3 (\gamma_1^2-2 \gamma_1) \Biggr]\,\,\,.
\nonumber
\end{equation} 
It is $\omega_0=-\frac{3}{2 k}$ and $\omega_1=\frac{1}{2k}$: 
since k is assumed to be $\geq 2$, it follows 
$\gamma _0=2-\frac{3}{2k}$, $(\gamma_0+1) \, mod \, 2=1-\frac{3}{2k}$, 
and therefore 
$[(\gamma_0+1) \, mod \, 2-1]^2=\frac{9}{4 k^2}$, 
$\gamma_1=\frac{1}{2k}$, hence 
$$B_2^{h.c.}\left(k,l=\frac{1}{2},T\right)=-\frac{\lambda_T^2}{8}-\frac{\lambda_T^2}{8}  \Biggl[ \frac{9}{4 k^2} + 3\left(\frac{1}{4 k^2}-\frac{1}{k}\right)
\Biggr]=-\frac{\lambda_T^2}{8} 
\left(1-\frac{3}{k}+\frac{3}{k^2} \right)\,\,\,.$$\\\\ 
$\star$ Case $l=1$ \ \ \ (with $k \geq 4$): 
\begin{equation} 
B_2^{h.c.}(k,l=1,T)=\frac{1}{(2l +1)^2} \sum^{2l}_{j=0} (2j+1) 
\left[\frac{1+(-1)^{j+2l}}{2} B^B_2(\omega_j,T) + 
\frac{1-(-1)^{j+2l}}{2}B^F_2(\omega_j,T)\right]= \nonumber\end{equation} 

$$=\frac{1}{9}\Biggl[ B^B_2(\omega_0,T)+3 B^F_2(\omega_1,T)+5 B^B_2(\omega_2,T)\Biggr]\,\,\,.$$ 
It is $\omega_0=-\frac{4}{k}$, 
$\omega_1=-\frac{2}{k}$, $\omega_2=\frac{2}{k}$ and the 
$\omega_j$'s are such that $\vert\omega_j\vert\leq 1$: therefore, 
by using (\ref{famousresult}), one has  
$$B_2^{h.c.}\left(k,l=1,T\right)=
\frac{\lambda_T^2}{36}\left[-1+4 \vert\omega_0\vert-2\omega_0^2+
3(1-2\omega_1^2)+5(-1+4 \vert\omega_2\vert-2\omega_2^2)\right]=$$ 
$$=\frac{\lambda_T^2}{36}\left[
-1+4 \cdot\frac{4}{k}-2\cdot \frac{16}{k^2}+
3 \left(1-2\cdot \frac{4}{k^2} \right)+
5 \left(-1+4\cdot \frac{2}{k}-2\cdot \frac{4}{k^2} \right)\right]=
\lambda_T^2\left( 
-\frac{1}{12}+\frac{14}{9 k}-\frac{8}{3 k^2} \right)\,\,\,.$$\\\\ 
$\star$ Case of the large-$k$ limit, 
with $\lim\limits_{l\rightarrow \infty, k \rightarrow \infty} 
\frac{l^2}{k}=a<\frac{1}{2}$: we limit ourselves to the case of even $2 l$ 
(the opposite one is similar). We define 
$j_{crit}$ as the maximum integer such that $\omega_j<0$, and $x_{crit}=\sqrt{2}l$: it can be verified that 
$\vert \omega_j\vert<1$ for all $j$. We have 
$$ B_2^{h.c.}\left(k,l,T\right)=\frac{1}{(2l+1)^2}\Biggl[\sum\limits_{j \text{ even}=0 }\limits^{2l} (2j+1)B_2^B (\omega_j,T)+ \sum\limits_{j \text{ odd}=1 }\limits^{2l-1} (2j+1)B_2^F (\omega_j,T)\Biggr]=$$ 
$$=\frac{\lambda_T^2/4}{(2l+1)^2}\Biggl[\sum\limits_{j \text{ even}=0 }\limits^{2l} (2j+1)(-1+4\vert\omega_j\vert-2\omega_j^2)+ \sum\limits_{j \text{ odd}=1 }\limits^{2l-1} (2j+1)(1-2\omega_j^2)\Biggr]= $$  
$$ = \frac{\lambda_T^2/4}{(2l+1)^2}\Biggl[\sum\limits_{j \text{ even}=0 }\limits^{j_{crit}} (2j+1)(-1-4\omega_j-2\omega_j^2)+\sum\limits_{j \text{ even}=j_{crit}+1 }\limits^{2l} (2j+1)(-1+4\omega_j-2\omega_j^2)$$
$$ + \sum\limits_{j \text{ odd}=1 }\limits^{2l-1} (2j+1)(1-2\omega_j^2)\Biggr]\simeq \frac{\lambda_T^2/4}{(2l+1)^2}\Biggl[\frac{1}{2}\int\limits_{0 }\limits^{x_{crit}} dx \, (2x+1)\Biggl(-1-4\frac{x^2-2l^2}{k}-2\frac{(x^2-2l^2)^2}{k^2}\Biggr)+$$
$$+\frac{1}{2}\int\limits_{x_{crit}}\limits^{2l} dx \, (2x+1)\Biggl(-1+4\frac{x^2-2l^2}{k}-2\frac{(x^2-2l^2)^2}{k^2}\Biggr) +\frac{1}{2} \int\limits_{0}\limits^{2l} dx\,(2x+1)\Biggl(1-2\frac{(x^2-2l^2)^2}{k^2}\Biggr)\Biggr]\simeq $$
$$\simeq \lambda_T^2\left(\frac{a}{2}-\frac{2}{3}a^2\right)\,\,\,.$$

\section{Comparison with Ref. \cite{Hagen96}} \label{hagen_appendix}
With the notation used in the main text, Eq. (2) of Ref. \cite{Hagen96} 
for NACS particles in the hard-core limit reads
\begin{equation} B_2^{h.c.}(\kappa,l, T) = \frac{1}{(2l +1)^2} \sum^{2l}_{j=0} (2j+1) 
\left[\frac{1+(-1)^{j+2l}}{2} B^B_2(\omega_j,T) + 
\frac{1-(-1)^{j+2l}}{2}B^F_2(\omega_j,T)\right]\,\,\,, \label{Hagen_app}\end{equation} 
where $\omega_j\equiv\frac{1}{4\pi\kappa} 
\left[j(j+1)-2l(l+1)\right]$ and 
$B^{B,F}_2(\omega,T)$ is given by \cite{Arovas85} 
\begin{equation} B^{B(F)}_2(\omega,T) = 
\frac{1}{4}\lambda^2_T \left\{ \begin{array}{ll} 
                 -1 + 4\delta - 2\delta^2, & \mbox{N even (odd)} \\ 
                 1 - 2\delta^2, & \mbox{N odd (even)} 
                 \end{array} 
                 \right. \label{famousresult_app}\end{equation} 
with $\omega=N+\delta$ and $N$ an integer such that $0 \leq \delta < 1$. 
Eq. (\ref{Hagen_app}) can be derived as in the following: with the notation 
$\gamma_j\equiv\omega_j \, mod \, 2$, using 
Eqs. (\ref{legamepartizioneviriale})-(\ref{limitenoninteragente})-(\ref{coeffviriale}) one has 
\begin{eqnarray} 
\nonumber B_2^{h.c.}(\kappa,l, T)=B_2^{(n.i.)}(l,T)-\frac{2\lambda_T^2}{(2l+1)^2}  
\left[Z^\prime_2 (\kappa, l, T) - Z^{\prime(n.i.)}_2 (l,T)\right]= 
\end{eqnarray} 
\begin{eqnarray}\nonumber 
=-\frac{1}{4}\frac{\lambda_T^2}{(2l+1)^2} 
\sum^{2l}_{j=0} (2j+1) \left[\frac{1+(-1)^{j+2l}}{2}+ \frac{1-(-1)^{j+2l}}{2}(-1)\right]+ 
\end{eqnarray} 
\begin{eqnarray}\nonumber 
-\frac{2 \lambda_T^2}{(2l+1)^2} 
\sum_{j=0}^{2l}(2j+1)\Biggl[\frac{1+(-1)^{j+2l}}{2} \frac{1}{4}(\gamma_j^2-2  
\gamma_j)+\frac{1-(-1)^{j+2l}}{2}\frac{1}{4}[(\gamma_j+1) \, mod \, 2-1]^2\Biggr]= 
\end{eqnarray} 
\begin{eqnarray}\nonumber 
=\frac{\lambda_T^2/4}{(2l+1)^2} 
\sum_{j=0}^{2l}(2j+1)\Biggl[\frac{1+(-1)^{j+2l}}{2}(-1+4 \gamma_j-2 \gamma_j^2)+\frac{1-(-1)^{j+2l}}{2}(1-2 [(\gamma_j+1) \, mod \, 2-1]^2)\Biggr]= 
\end{eqnarray} 
\begin{displaymath}\nonumber 
=\frac{1}{4}\frac{\lambda_T^2}{(2l+1)^2} 
\sum_{j=0}^{2l}(2j+1)\Biggl[\frac{1+(-1)^{j+2l}}{2}\left\{ \begin{array}{ll} 
                 -1 + 4\delta_j - 2\delta_j^2, & \mbox{$N_j$ even} \\ 
                 1 - 2\delta_j^2, & \mbox{$N_j$ odd} 
                 \end{array} 
                 \right\} + 
\end{displaymath} 
\begin{displaymath}\nonumber 
+\frac{1-(-1)^{j+2l}}{2}\left\{ \begin{array}{ll} 
                 1 - 2\delta_j^2, & \mbox{$N_j$ even} \\ 
                 -1 + 4\delta_j - 2\delta_j^2, & \mbox{$N_j$ odd} 
                 \end{array} 
                 \right\}\Biggr]= 
\end{displaymath} 
\begin{eqnarray} \nonumber 
=\frac{1}{(2l +1)^2} \sum^{2l}_{j=0} (2j+1) 
\left[\frac{1+(-1)^{j+2l}}{2} B^B_2(\omega_j,T) + 
\frac{1-(-1)^{j+2l}}{2}B^F_2(\omega_j,T)\right]\,\,\,, 
\end{eqnarray} 
that is nothing else than the (\ref{Hagen_app}) itself. 

Notice that Eq. (3) of \cite{Hagen96} should be replaced with 
\begin{equation}\frac{4}{\lambda_T^2} B_2^{h.c.}(\alpha,T)=\frac{1}{(2l+1)^2}\sum_{j=0}^{2 l} (2j+1)[\mp(-1)^{j+2l}]+\frac{2}{(2 l+1)^2}\sum_{j=0}^{2 l} (2j+1)\delta_j[1\pm(-1)^{j+2l}-\delta_j]\,\,\,,\label{trecorretta}\end{equation} 
where the upper and lower signs refer to the cases of even and odd $N_j$'s: 
Eq. (\ref{trecorretta}) is equivalent to our formula (\ref{coeffviriale}) and 
to Eq. (\ref{Hagen_app}).

%In \cite{Hagen96} it is stated that 
%for $\alpha\equiv\frac{1}{4 \pi \kappa}\rightarrow 0$ 
%Eq. (30) of \cite{Lee95} would predict that the virial coefficient 
%becomes the free bosonic result $B_2^B(T)=-\lambda_T^2 /4$. 
%This appears to be not the case: indeed, 
%although the $\omega_j$'s tend to $0$ in this "free" limit, 
%the same does not stand for their respective integer parts $\delta_j$ 
%appearing in this formula. 
We discuss now the limit 
$\alpha\equiv\frac{1}{4 \pi \kappa}\rightarrow 0$ : we observe 
that by a careful inspection it is possible to conclude that 
Eq. (30) of \cite{Lee95} and Eq. (3) of \cite{Hagen96} 
do not tend  in this limit to the correct value $B_2^{(n.i.)}(l,T)$ given 
in Eq. (\ref{deltabidue}). 
However, the manipulation of the corrected version 
(\ref{trecorretta}) above presented reproduces (as expected) 
the value $B_2^{h.c.} \rightarrow -\frac{\lambda_T^2}{4}\frac{1}{2 l +1}$ 
for $\alpha \rightarrow 0$. 
Indeed for vanishing coupling constant $\alpha$, using 
the same convention used above (upper and lower choices 
referring to the cases of even and odd $N_j$ respectively), one has
$$\frac{1}{4 \pi \kappa}\rightarrow 0\Rightarrow\frac{1}{(2l+1)^2}\sum_{j=0}^{2 l} (2j+1)[\mp(-1)^{j+2l}]+\frac{2}{(2 l+1)^2}\sum_{j=0}^{2 l} (2j+1)\delta_j[1\pm(-1)^{j+2l}-\delta_j] \rightarrow $$ 
$$ \frac{2}{(2 l+1)^2}\sum_{j=0}^{2 l} (2j+1)(\frac{1}{2} \mp \frac{1}{2} )[1\pm(-1)^{j+2l}-(\frac{1}{2} \mp \frac{1}{2} )] + \frac{1}{(2l+1)^2}\sum_{j=0}^{2 l} (2j+1)[\mp(-1)^{j+2l}]=$$ 
$$
=-\frac{\sum_{j=0}^{2 l}(2j+1)(-1)^{j+2l}}{(2l+1)^2}=-\frac{1}{2l+1}\,\,\,,$$
as it might.

\section{Spectrum in the soft-core non-Abelian case}\label{spectrum_sc}
In Subsection \ref{generalsoftcorecase} 
we stated that the spectrum of the multi-component 
projected Hamiltonian operator $H_j'$ 
can be represented in general as the union of the $(2j+1)$ spectra 
of the corresponding scalar Schr\"odinger operators. That follows from the following remark: the non-Abelian generalization of the soft-core expression (\ref{sol}) is 
Eq. (\ref{nonabelianradialsoftcore}). By denoting 
$$A(r)\equiv\frac{1}{M}\left[-\frac{1}{r}\frac{d}{dr} 
r \frac{d}{dr}+\frac{(n+\alpha)^2}{r^2}\right]\,\,\,,$$ the non-Abelian generalization of (\ref{eqradiale}) is the $(2l+1)^2-$dimensional matricial equation 
\begin{equation}
\label{equazionematricialedisaccoppiata} 
\left( 
\begin{array}{ccccc} 
A(r) & 0 & 0 & \cdots & \\ 
0 & A(r) & 0 & \cdots & \\ 
0 & 0 &  A(r) & \cdots & \\ 
\cdots & \cdots & \cdots & \cdots & 
\end{array} 
\right) 
\left( 
\begin{array}{c} 
R_0^{0,0} \\ 
R_0^{1,-1} \\ 
R_0^{1,0}  \\ 
\cdots \\ 
\end{array} 
\right) 
=E 
\left(
\begin{array}{c} 
R_0^{0,0} \\ 
R_0^{1,-1} \\ 
R_0^{1,0}  \\ 
\cdots \\ 
\end{array}
\right) 
=\frac{k^2}{M} 
\left(
\begin{array}{c} 
R_0^{0,0} \\ 
R_0^{1,-1} \\ 
R_0^{1,0}  \\ 
\cdots \\ 
\end{array}
\right) \,\,\,.
\end{equation}

The hard-disk regularization $R_0^{j,j_z}(r=R)=0$ for all $j,j_z$ 
discretizes the energy spectrum; let then $Sp_{j,j_z}(R)$ be the 
(discretized) spectrum of the component equation 
$A(r)R_0^{j,j_z}=ER_0^{j,j_z}$ restricted over the domain 
in which $R_0^{j,j_z}$ has hard-core parameter $\varepsilon_{j,j_z}$, 
and $Sp(R)$ be the (discretized) spectrum of 
Eq. (\ref{equazionematricialedisaccoppiata}) 
in the domain in which any component $R_0^{j',j'_z}$ 
has the respective assigned hard-core parameter $\varepsilon_{j',j'_z}$. 
If $E_{j',j'_z}\in Sp_{j',j'_z}(R)$ for some $(j',j'_z)$, then 
also $E_{j',j'_z}\in Sp(R)$, because   
\begin{equation} 
\left( 
\begin{array}{ccccc} 
A(r) & 0 & 0 & \cdots & \cdots \\ 
0 & A(r) & 0 & \cdots & \cdots \\ 
0 & 0 &  A(r) & \cdots & \cdots \\ 
\cdots & \cdots & \cdots & \cdots & \cdots\\
\cdots & \cdots & \cdots & \cdots & \cdots
\end{array} 
\right) 
\left( 
\begin{array}{c} 
0 \\ 
\cdots \\ 
R_0^{j',j'_{z}}\\ 
0\\ 
\cdots  
\end{array}
\right)  
=E_{j',j'_z}
\left( 
\begin{array}{c} 
0 \\ 
\cdots \\ 
R_0^{j',j'_z}\\ 
0\\ 
\cdots  
\end{array} 
\right)\,\,\,, 
\end{equation} 
and all the null components of the vector trivially 
fulfill whichever hard-core conditions, 
in particular the assigned sequence 
$\{\varepsilon_{j,j_z}\}\in \{[0,\infty)\}^{(2l+1)^2}.$ 
In conclusion, the spectrum in the non-Abelian case 
can be written as the above-mentioned union of spectra, 
which will automatically include all the possible relevant energy 
degenerations to be considered in the partition function 
for the computation of the virial coefficients.

\end{document}